\documentclass[prd,showpacs,groupedaddress,superscriptaddress,amsmath,amssymb]{revtex4-2} 

\usepackage{hhline}
\usepackage{slashed}
\usepackage{graphicx}
\usepackage{graphics}
\usepackage{dcolumn}
\usepackage{bm}
\usepackage{amssymb}
\usepackage{enumerate}
\usepackage{lineno}
\usepackage{multirow}
\usepackage{orcidlink}
\usepackage{slashed}

\newcommand\ee{e^+e^-}

\newcommand\Zpr{Z^\prime}

\begin{document}


\title{ \bf Development of the fully Geant4 compatible package for the simulation of Dark Matter in fixed target experiments}



\author{B.~Banto Oberhauser\orcidlink{0009-0006-4795-1008}}
\affiliation{ETH Z\"urich, Institute for Particle Physics and Astrophysics, CH-8093 Z\"urich, Switzerland}

\author{P.~Bisio\orcidlink{0009-0006-8677-7495}}
\affiliation{INFN, Sezione di Genova, 16147 Genova, Italia}
\affiliation{Universit\'a degli Studi di Genova, 16126 Genova, Italia}

\author{A.~Celentano\orcidlink{0000-0002-7104-2983}}
\affiliation{INFN, Sezione di Genova, 16147 Genova, Italia}

\author{E.~Depero\orcidlink{0000-0003-2239-1746}}
\affiliation{ETH Z\"urich, Institute for Particle Physics and Astrophysics, CH-8093 Z\"urich, Switzerland}

\author{R.~R.~Dusaev\orcidlink{0000-0002-6147-8038}}
\affiliation{Tomsk Polytechnic University, 634050 Tomsk, Russia}

\author{D.~V.~Kirpichnikov\orcidlink{0000-0002-7177-077X}}
\affiliation{Institute for Nuclear Research, 117312 Moscow, Russia}

\author{M.~M.~Kirsanov\orcidlink{0000-0002-8879-6538}}
\email[\textbf{e-mail}: ]{mikhail.kirsanov@cern.ch}
\affiliation{Institute for Nuclear Research, 117312 Moscow, Russia}

\author{N.~V.~Krasnikov\orcidlink{0000-0002-8717-6492}}
\affiliation{Institute for Nuclear Research, 117312 Moscow, Russia}
\affiliation{Joint Institute for Nuclear Research, 141980 Dubna, Russia}

\author{A.~Marini\orcidlink{0000-0002-6778-2161}}
\affiliation{INFN, Sezione di Genova, 16147 Genova, Italia}

\author{L.~Marsicano\orcidlink{0000-0002-8931-7498}}
\affiliation{INFN, Sezione di Genova, 16147 Genova, Italia}

\author{L.~Molina-Bueno\orcidlink{0000-0001-9720-9764}}
\affiliation{Instituto de Física Corpuscular, Universidad de Valencia and CSIC, Carrer del Catedrátic José Beltrán Martinez, 2, 46980 Paterna, Valencia, Spain}

\author{M.~Mongillo\orcidlink{0009-0000-7331-4076}}
\affiliation{ETH Z\"urich, Institute for Particle Physics and Astrophysics, CH-8093 Z\"urich, Switzerland}

\author{D.~Shchukin\orcidlink{0009-0007-5508-3615}}
\affiliation{P.N. Lebedev Physical Institute, Moscow, Russia, 119991 Moscow, Russia}

\author{H.~Sieber\orcidlink{0000-0003-1476-4258}}
\affiliation{ETH Z\"urich, Institute for Particle Physics and Astrophysics, CH-8093 Z\"urich, Switzerland}

\author{I.~V.~Voronchikhin\orcidlink{0000-0003-3037-636X}}
\affiliation{Tomsk Polytechnic University, 634050 Tomsk, Russia}


\date{\today}

\begin{abstract}
 The search for new comparably light (well below the electroweak scale) feebly interacting particles is an exciting possibility
to explain some mysterious phenomena in physics, among them the origin of Dark Matter. The sensitivity study through detailed
simulation of projected experiments is a key point in estimating their potential for discovery.

 Several years ago we created the DMG4 package for the simulation of DM (Dark Matter) particles in fixed target experiments.
The natural approach is to integrate this simulation into the same program that performs the full simulation of particles in the
experiment setup. The Geant4 toolkit framework was chosen as the most popular and versatile solution nowadays.

 The simulation of DM particles production by this package accommodates several possible scenarios, employing electron,
muon or photon beams and involving various mediators, such as vector, axial vector, scalar, pseudoscalar, or spin 2 particles.
The bremsstrahlung, annihilation or Primakoff processes can be simulated.

 The package DMG4 contains a subpackage DarkMatter with cross section methods weakly connected to Geant4. It can be used in different
frameworks.

 In this paper, we present the latest developments of the package, such as extending the list of possible mediator particle types,
refining formulas for the simulation and extending the mediator mass range. The user interface is also made more flexible and convenient.

 In this work, we also demonstrate the usage of the package, the improvements in the simulation accuracy and some cross check validations.

\end{abstract}

\maketitle
\newpage

\section*{New version program summary}

\emph{Program title:} DMG4 \\
\emph{CPC Library link to program files:} \\
\emph{Code Ocean capsule:} \\
\emph{Licensing provisions:} GNU General Public License 3 (GPL) \\
\emph{Programming language:} c++ \\
\emph{Nature of problem:} For the simulation of Dark Matter production processes in fixed target experiments a code that can be easily
 integrated in programs for the full simulation of experimental setup is needed. \\
\emph{Solution method:} A fully Geant4 compatible DM simulation package DMG4 was presented in 2020. We present numerous further
 developments of this package.

\section{Introduction}

 The \emph{DMG4} package is designed for simulating the production of feebly interacting particles 
beyond the Standard Model (SM) present in many models trying to explain Dark Matter and various anomalies observed in
particle physics. It deals with particles of masses well below the electroweak scale, extending down to $\sim 1$ keV. The experimental technique 
involving a fixed thick target is very popular in the searches for these particles
\cite{Bjorken:2009mm,Izaguirre:2013uxa,Batell:2009di,Izaguirre:2014bca}. It is convenient to simulate
their production in the same program that is used for the simulation of the whole setup. For this reason, the DMG4 package by construction
is made fully compatible with the Geant4 toolkit \cite{Agostinelli:2002hh}, which is a foundation for the majority of such programs.
The initial description of the DMG4 package is presented in ref. \cite{Bondi:2021nfp}. In this paper we describe its further
developments.

 While the most popular DM model introduces a vector particle, ``dark photon''~\cite{Holdom:1985ag} as a mediator between SM and DM particles,
different quantum numbers and couplings of such mediators are possible \cite{Battaglieri:2017aum,Beacham:2019nyx}. We tried to provide
the possibility of simulation for as wide as possible class of models that can be relevant for the searches for BSM (beyond SM) particles
at accelerators with moderate beam energies $\sim\mathcal{O}(10-100)$ GeV. The recent extensions we have implemented are presented in this article.

Several relevant production processes involve $2 \rightarrow 3$ interactions on a heavy nucleus, for which is difficult to derive exactly all 
formulas needed for efficient simulation. Various approximations are often used, introducing deviations from results obtained at
exact tree level (ETL). The latter results serve as reference, but cannot be used in the codes that perform event-by-event sampling.
We performed a number of studies and developments in order to improve the accuracy of simulation, which we present in this article.

 The DM mediators can be produced through annihilation processes, either of electromagnetic shower positrons or beam positrons with
target electrons. The corresponding exact formulas were derived and used in the package. However, for such processes, there were
other implementation problems that we have recently addressed. We present the corresponding solutions here.

\section{DMG4 package structure}

Structurally, the DMG4 package consists of two main components: a set of lower-level
routines responsible for DM-related calculus, and a layer of Geant4 interfaces
realization. This way various numerical tests and model benchmarking can be
implemented using only the numerical routines part, before it gets embedded
into Geant4 pipeline.

As for numerical routines, every process subclasses common parent root
\texttt{DarkMatter} (being an \emph{abstract product} in terms of
\emph{factory pattern}) facilitating common requirements for dark matter models
in terms of mean free path, spatial dependencies on cross sections, decay width,
etc. Subclasses implementing a particular model (\emph{concrete products})
implement its interface (\emph{DM production process}) by overriding
certain virtual methods of \texttt{DarkMatter} base class.

\begin{figure*}[h]
\begin{center}
\includegraphics[width=0.75\textwidth]{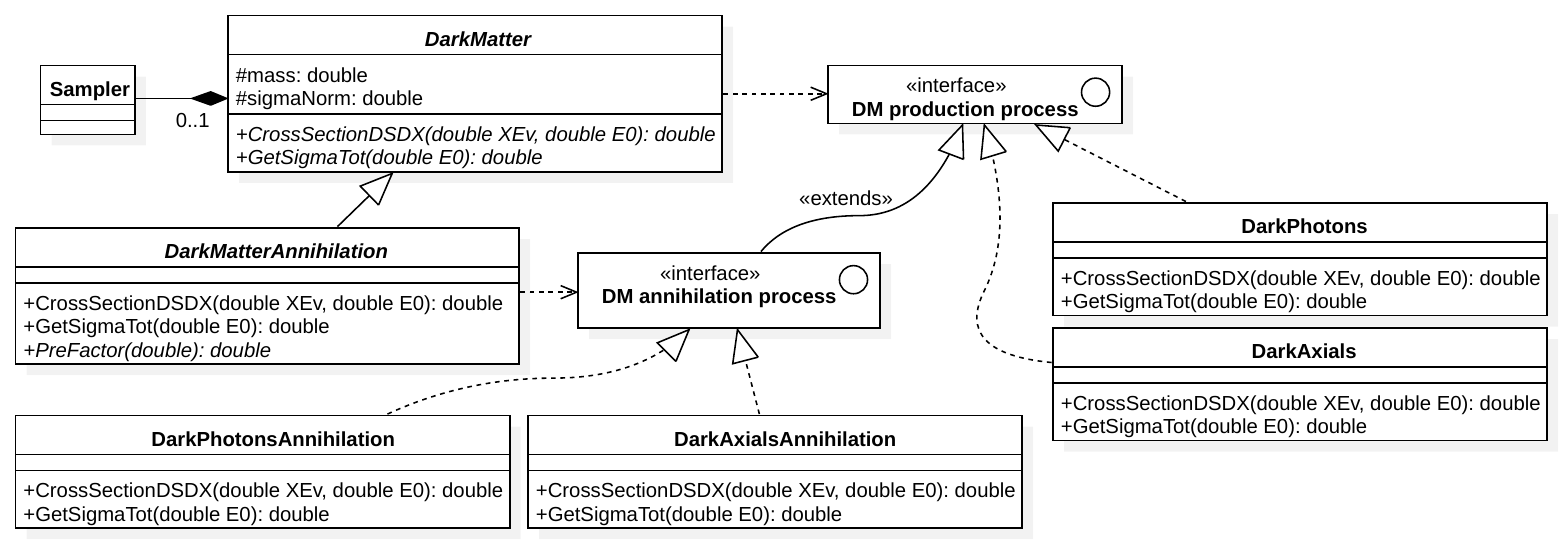}
    \caption {Classes of the main DMG4 interfaces shown as
    UML~diagram~\cite{iso_uml2_standard}. For brevity, in bremsstrahlung part only dark
    photons and dark axials classes are shown.
\label{fig:ClassesDiagram}}
\end{center}
\end{figure*}

The base \texttt{DarkMatter} class is extended by
the \emph{DM annihilation process} interface within \texttt{DarkMatterAnnihilation}.
The \emph{DM annihilation process} appends contract with routines yielding certain
resonant-specific quantities (additional yield factor and angular
distribution PDF). The diagram~\ref{fig:ClassesDiagram} illustrates relationships
between interfaces and corresponding implementation classes with dark axial and
dark photon modules as examples of concrete products. A particular sampler
instance can be associated with physics model at the most abstract level.

\begin{figure*}[h]
\begin{center}
\includegraphics[width=0.75\textwidth]{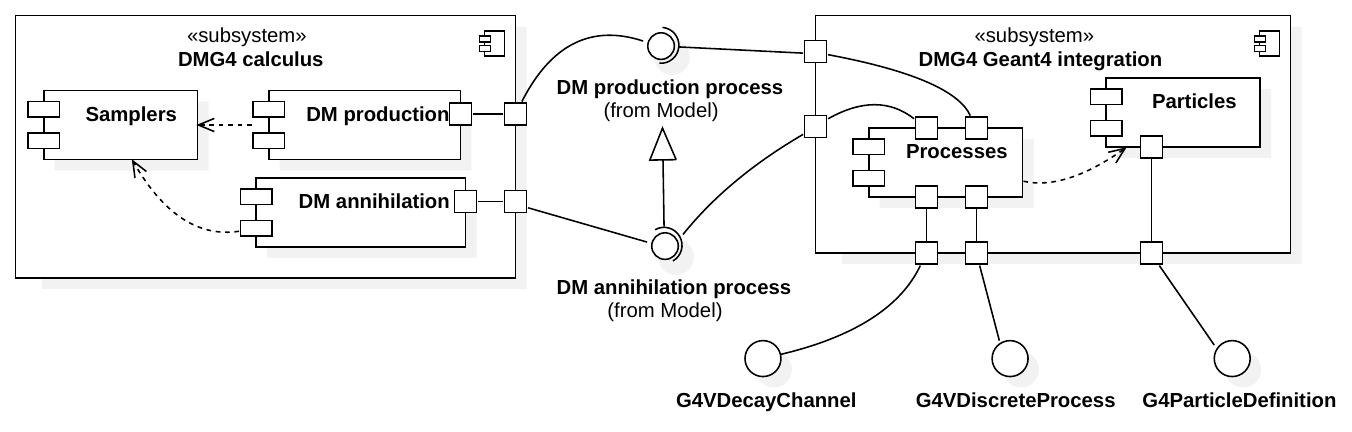}
    \caption {UML~Components~diagram~\cite{iso_uml2_standard} of DMG4 package exposing Geant4-compatible facade.
\label{fig:ComponentDiagram}}
\end{center}
\end{figure*}

Component diagram \ref{fig:ComponentDiagram} provides a wider view on how
particular numerical routines are exposed in terms of Geant4 API (namely,
\texttt{G4ParticleDefinition} and \texttt{G4VDiscreteProcess}). Corresponding
interfaces are implemented within the interim layer composed of a set of
processes and related particles while the cross-section calculus is expressed
by means of internal classes.

The DM sector particles introduced in the package are listed in Table~\ref{table:dmparticles}.
The PDG codes are ascribed according to the slightly extended rules in~\cite{PDG_particles}. As compared to early versions,
we reduced the number of mediator particle classes, which was possible due to usage of a new class \emph{DarkMatterParametersFactory}.
Using this class it is more convenient to specify the decay modes and some other parameters of particles. On the other hand,
we added several DM particles of models with semivisible decay modes. Note that in many configurations we assume the
mediators to be stable, although in full models they decay into stable DM particles. But for the missing energy signature simulation
it is unimportant as the energy carried away is the same.


\begin{table}[h]
\caption{DM particles defined in the package DMG4}
\begin{center}
\begin{tabular}{|c|c|c|c|c|c|c|c|}
\hline
Name                      &  PDG ID      & emitted by & spin & parity & Model                  & stable? & decay                      \\
\hline
DMParticleAPrime          &  5500022     & $e^+,e^-$  &  1   &  1     &                        & true    & -                          \\
                          &  5500122     &            &      &        & Visible X              & false   & $e^+e^-$                   \\
                          &  5500222     &            &      &        & B - L                  & false   & $e^+e^-$, $\mu^+\mu^-$ etc.\\
                          &  5500322     &            &      &        & Inelastic DM           & false   & $\chi_1, \chi_2$           \\
\hline
DMParticleScalar          &  5400022     & $e^+,e^-$  &  0   &  1     &                        & true    & -                          \\
                          &  5400022     &            &      &        & Visible X              & false   & $e^+e^-$                   \\
\hline
DMParticlePseudoScalar    &  5410022     & $e^+,e^-$  &  0   & -1     &                        & true    & -                          \\
                          &  5410122     &            &      &        & Visible X              & false   & $e^+e^-$                   \\
\hline
DMParticleAxial           &  5510022     & $e^+,e^-$  &  1   & -1     &                        & true    & -                          \\
                          &  5510122     &            &      &        &                        &         & $e^+e^-$                   \\
\hline
DMParticleZPrime          &  5500023     & $\mu$      &  1   &  1     & $L_{\mu}-L_{\tau}$     & true    & -                          \\
                          &  5500023     & $\mu$      &      &        & $L_{\mu}-L_{\tau}$     & false   & $\nu \nu$, $\mu^+\mu^-$    \\
\hline
DMParticleALP             &  5300122     & $\gamma$   &  0   & -1     &                        & false   & $\gamma \gamma$            \\
\hline
DMParticleChi1            &  5200014     & mediator   & 1/2  & 1      & Inelastic DM           & true    & -                          \\
\hline
DMParticleChi2            &  5200013     & mediator   & 1/2  & 1      & Inelastic DM           & false   & $\chi_1 e^+ e^-$           \\
\hline
DMParticleChi             &  5200012     & mediator   & 1/2  & 1      &                        & true    & -                          \\
\hline
\end{tabular}
\end{center}
\label{table:dmparticles}
\end{table}

 The dark sector particles are assumed to be stable as they are anyway undetectable. This does not apply to the $\chi_2$ appearing
in models of inelastic Dark Matter\cite{Tucker-Smith:2001myb}. For this particle, the 3-body decay is simulated
using the matrix element sampling implemented in the new \texttt{G4iDM3bodyDecayChannel} class, as explained in Sec.\ref{sec:idm}.

The current version of DMG4 package contains the following processes of DM production:

\begin{itemize}
\item
 Bremsstrahlung-like process of the type $b N \to b N X$, where $b$ is a projectile (can be $e^-, e^+, \mu^-, \mu^+$),
and $X$ is a DM mediator. Apart from the mediators listed in the table, we added a process with a spin 2 particle without adding
a corresponding particle class. It means it is possible to simulate the invisibly decaying spin 2 mediator.

\item
 Primakoff process of photon conversion $\gamma N \to a N$, where $a$ is an axion-like particle (ALP)~\cite{Dusaev:2020gxi}

\item
 Resonant in-flight positron annihilation on atomic electrons $e^+ e^- \rightarrow X \rightarrow \chi \chi$, where $\chi$ is a dark
matter mediator decay product~\cite{Marsicano:2018glj}.

\end{itemize}

 The physics for a simulation run is configured in the function \texttt{DarkMatterPhysicsConfigure} called from the constructor
of the steering class \texttt{DarkMatterPhysics}. The configuration is done using the singleton class \emph{DarkMatterParametersFactory}.
A user specifies a number of parameters, then the steering class performs the needed actions:
\begin{itemize}
\item
Creates an instance of one of the concrete classes corresponding to the needed process and derived from the base class \texttt{DarkMatter},
for example \texttt{DarkPhotons}
\item
Instantiates and registers the needed particles and processes provided by the DMG4 package in terms of the native Geant4 API.
\end{itemize}
 The following main parameters are to be specified: 1. The process code; 2. The mixing (or coupling) parameter $\epsilon$; 3. The mediator mass;
4. The decay mode; 5. The cut-off minimal energy of particles that can initiate the processes of DM production. The latter is needed
to avoid the simulation of very soft DM particles that are anyway undetectable. In the new versions, this parameter also specifies, for some
processes or for a part of the mediator mass range, a minimal mediator energy that is allowed to be generated.

\section{Package DarkMatter}\label{sec:dark-matter}

 This class, which we try to keep as weakly connected to Geant4 as possible, contains the methods that return total and
differential cross sections, first of all for the bremsstrahlung processes. For these processes the cross sections can be
calculated numerically at Exact Tree Level (ETL) or using the formulas of Weizsaker-Williams (WW)
and improved Weizsaker-Williams (IWW) approximations \cite{Bjorken:2009mm}.

 As explained in \cite{Bondi:2021nfp}, for the electron beam it contains the tabulated K-factors that correct the total cross
sections obtained in IWW approximations for the electron beam to the values calculated at ETL. As compared to
the previous version, we extended the corresponding tables down to the mediator masses of 1 keV and, for some classes, up to 3 GeV.

 In addition to this, we also tabulated the differential cross sections $d\sigma/dx$ for the electron beam and mediator masses below 1 MeV,
where it can have non-trivial shape, significantly different from the sharply peaked one at around $x=1$ for heavier
mediators.

 The Lagrangians and initial cross section formulas used in the package can be found in the previous paper \cite{Bondi:2021nfp}.
The Lagrangian for the new model with spin 2 mediator looks like this \cite{Voronchikhin_2022,lee:2014GMDM,Kang:2020-LGMDM}:
\begin{align}
&    \mathcal{L}^{\rm G}_{\rm eff} 
\supset 
 -
    \frac{i c^{\rm G}_{ee}}{2\Lambda} G^{\mu \nu}
    \left(
            \overline{e} \gamma_{\mu} \overleftrightarrow{D}_{\nu} e
        -   \eta_{\mu \nu} \overline{e} \gamma_{\rho} \overleftrightarrow{D}^{\rho} e
    \right) \label{BechmarkLagrOftheModel1}  
\\& +  \frac{c^{\rm G}_{\gamma \gamma}}{\Lambda} G^{\mu \nu} 
    \left(
        \frac{1}{4} \eta_{\mu \nu} F_{\lambda \rho} F^{\lambda \rho} 
    +   F_{\mu \lambda} F^{\lambda}_{ \nu}
    \right) +
     \frac{c^{\rm G}_{\rm DM}}{\Lambda} G^{\mu \nu} T^{\rm DM}_{\mu \nu}, 
     \nonumber
\end{align}
where $e$ is the label of the SM electron, $F_{\mu\nu} = \partial_\mu A_\nu - \partial_\nu A_\mu$ is a 
stress tensor of the SM photon field $A_\mu$, $D_\mu = \partial_\mu - i e A_\mu$ is a covariant derivative of the $U(1)$ gauge field,  
and $\Lambda$ is the dimensional parameter for spin-2 interactions, that is associated with the scale of new physics;
$c_{ee}^{\rm G}$ and  $c_{\gamma \gamma}^{\rm G}$ are dimensionless couplings for  the  
electron and photon respectively. We choose the universal coupling $c^{\rm G}_{ee} = c^{\rm G}_{\gamma \gamma}$ throughout the paper.

 Finally, since the early versions of the package, significant progress has been made in the cross sections for the case of a muon beam and
in the annihilation processes. These, alongside other developments, are described in the following sections.

\section{Inclusion of inelastic Dark Matter models}\label{sec:idm}

In multi-generational dark sector models, such as scenarios involving inelastic Dark Matter, the dark photon exhibits off-diagonal
couplings to a pair of particles, denoted as $\chi_1$ and $\chi_2$, with different masses. The heavier particle, $\chi_2$, is unstable
and decays via the channel $\chi_2\rightarrow\chi_1e^-e^+$. In such cases, the dark photon decay cascade leads to both visible
and invisible final states and is consequently referred to as semivisible $A^\prime$.

To enable the simulation of these next-to-minimal models, two new particles, \texttt{DMParticleChi1} and \texttt{DMParticleChi2},
 have been added to the DMG4 package. Notably, two distinct semivisible models are now included: inelastic Dark Matter
(iDM)~\cite{Tucker-Smith:2001myb} and inelastic Dirac Dark Matter (i2DM)~\cite{Filimonova_2022}. The decay widths for these
models are taken from Refs.~\cite{Mohlabeng:2019vrz, NA64:2021acr} and Ref.\cite{Filimonova_2022} respectively.

\begin{figure}[h]
    \centering
    \includegraphics[width=0.6\textwidth]{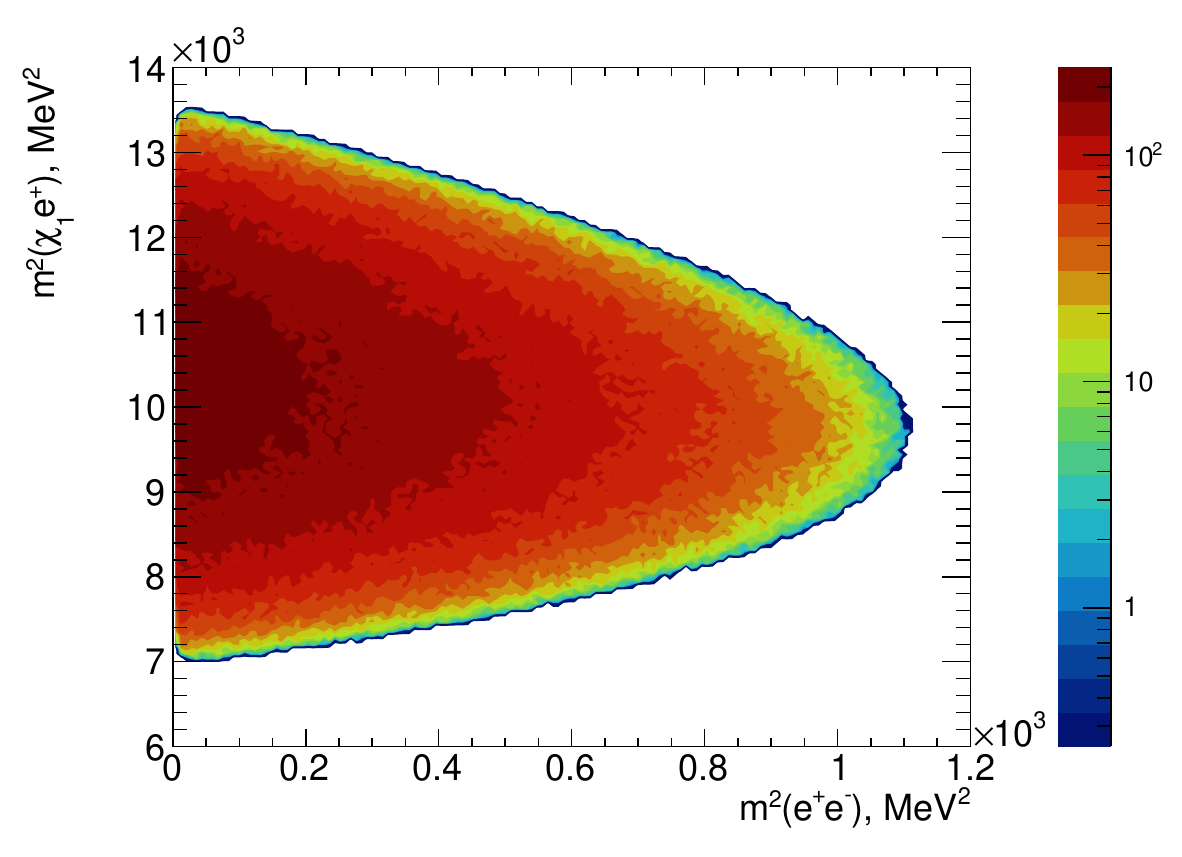}
    \caption{Dalitz plot for $\chi_2\rightarrow\chi_1e^+e^-$ obtained through the matrix element sampling from the method
             \texttt{G4iDM3bodyDecayChannel} for $m_{A^\prime}=0.25$ GeV, $m_{A^\prime}=3m_{\chi_1}$ and $f=0.4$.}
    \label{fig:3bodydecay}
\end{figure}

To account for the specific 3-body decay process $\chi_2\rightarrow\chi_1e^-e^+$, the DMG4 class \texttt{G4iDM3bodyDecayChannel}
has been implemented. This class, derived from \texttt{G4VDecayChannel} and based on \texttt{G4DalitzDecayChannel}, performs
the sampling of the spin-averaged matrix element, as illustrated in Fig.~\ref{fig:3bodydecay}. The kinematics of the final state
particles is subsequently calculated based on the obtained squared invariant masses of the $e^+e^-$ and $\chi_1e^+$ systems.

\section{Angular sampling of ALP production via Primakoff effect}
\begin{figure}[h]
    \centering
    \includegraphics[width=0.8\textwidth]{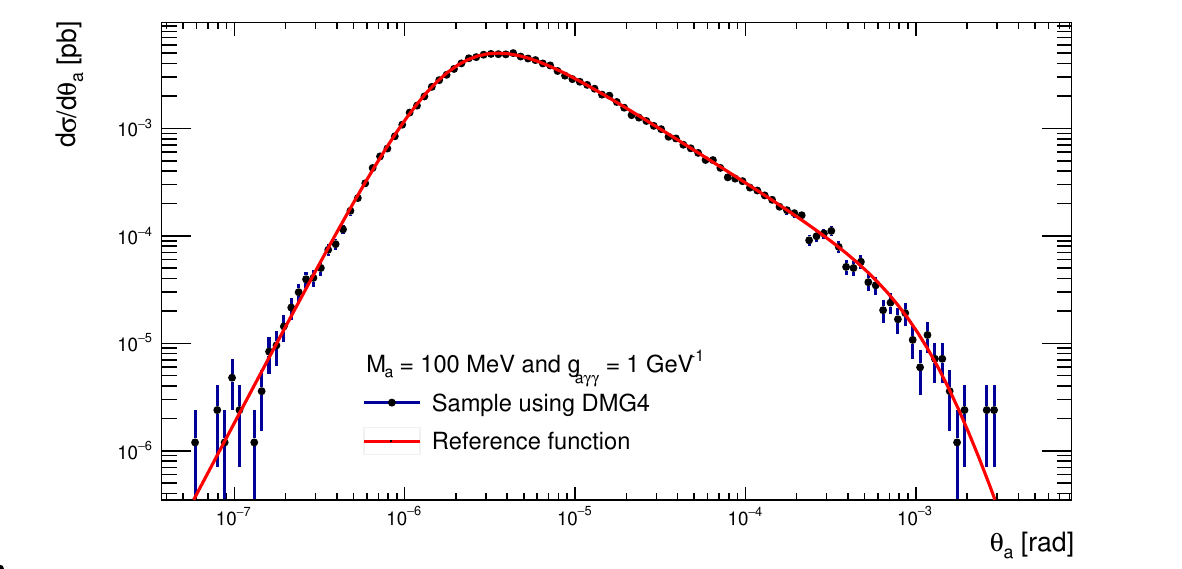}
    \caption{Comparison between results obtained using the sampling from the method
    \texttt{SimulateEmissionWithAngle3} and the reference function for the differential cross-section $d\sigma/d\theta_{a}$ of the Primakoff 
process $\gamma N \rightarrow a N$ \cite{Dusaev:2020gxi}. The cross-section was calculated for an incoming photon energy $E_{\gamma} = 50$~GeV.}
    \label{fig:dsdtheta-ALP}
\end{figure}

In order to simulate the final state kinematics of the ALP produced in the Primakoff process  $\gamma N \rightarrow a N$,
the sampling of the emission angle was implemented in DMG4. The differential cross section with respect to the angle of ALP emission
is obtained in the limit when $m_a \ll E_a$ and $\theta_a \ll 1$ and is used in the sampling scheme, as presented in our
previous work \cite{Dusaev:2020gxi}. In this approximation the energy of the emitted ALP $E_{a}$ can be calculated
based on the angle $\theta_{a}$. The results of the angular sampling are displayed in Fig. \ref{fig:dsdtheta-ALP} and compared
to the reference distribution function, which is a single differential cross section.

\section{Developments in muon-philic processes and cross sections implementations}

In this section, we describe the recent developments in the simulation of light dark matter production in bremsstrahlung-like
reactions of muons on nuclei, $\mu N\rightarrow\mu NX$. In particular, we describe the extension of muon-philic models to 
different mediator types and the improvements in the underlying production cross sections implementation.

\subsection{Broadening the class of mediators}

In our previous work \cite{Bondi:2021nfp}, the emission of a light mediator from muons was associated with the $L_\mu-L_\tau$ gauge
vector boson, $X=Z'$, \cite{He:1990pn,He:1991qd,Foot:1994vd,Altmannshofer:2016jzy,Kile:2014jea,Park:2015gdo}. Within this model, the
illustrative Lagrangian of the theory is given by
\begin{equation}
    \mathcal{L}\supset\mathcal{L}_{SM}-\frac{1}{4}F_{\mu\nu}^\prime F^{\mu\nu\prime}+\frac{m_{Z^\prime}^2}{2}Z_\mu^\prime Z^{\mu\prime}-g_{Z^\prime}Z_\alpha^{\prime}\big(\bar{\mu}\gamma^\alpha\mu-\bar{\tau}\gamma^\alpha\tau
        +\bar{\nu}_\mu\gamma^\alpha P_L\nu_\mu
        -\bar{\nu}_\tau\gamma^\alpha P_L\nu_\tau\big),
\end{equation}
where $F_{\mu\nu}^\prime$ is the field strength tensor associated with the $Z_\mu^\prime$ field, $m_{Z'}$ the mass of the gauge boson,
$g_{Z'}=\epsilon_{Z'}e$ the coupling to SM leptons and $P_{L}$ the left-handed chiral projection operator. The decay width associated
with the \emph{vanilla} model is given by the purely invisible channel to SM neutrinos, $Z'\rightarrow\bar{\nu}\nu$, such that
\begin{equation}
    \Gamma_{Z'\rightarrow\bar{\nu}\nu}=\frac{\alpha_{Z'}m_{Z'}}{3},
\end{equation}
where $\alpha_{Z'}=g_{Z'}^2/(4\pi)$.\\ \indent
The addition to the aforementioned model, the formulas for other types of light muon-philic mediators is provided, namely for scalar and 
pseudoscalar particles, respectively $X=S$ and $X=P$, with corresponding simplified Lagrangians
\begin{equation}
	\label{eq:lagrangian-scalar-muon}
	\mathcal{L} \supset \mathcal{L}_{SM} + \frac{1}{2} (\partial_\mu S)^2 - \frac{1}{2} m_S^2 S^2 +
g_S  S \overline{\mu} \mu,
\end{equation}
\begin{equation}
	\label{eq:lagrangian-pseudoscalar-muon}
	\mathcal{L} \supset \mathcal{L}_{SM} + \frac{1}{2} (\partial_\mu P)^2 - \frac{1}{2} m_P^2 P^2
+ i   g_{P}    P \overline{\mu} \gamma_5 \mu,
\end{equation}
where it is assumed that the mediators do not need to couple to neutrinos, have masses $m_S$ and $m_P$ and muon-specific
couplings $g_S$ and $g_P$.

\subsection{Calculations of the production cross sections\label{subsec:muon-cross-sections}}

\begin{figure}[h]
    \centering
    \includegraphics[width=0.9\textwidth]{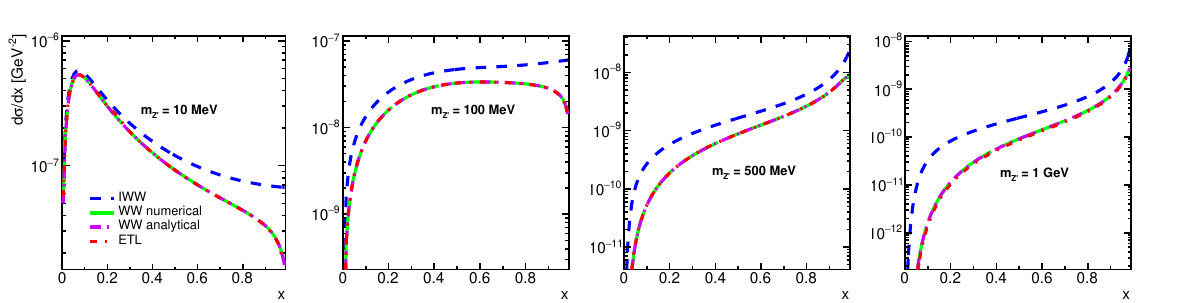}
    \hspace{1mm}
    \includegraphics[width=0.9\textwidth]{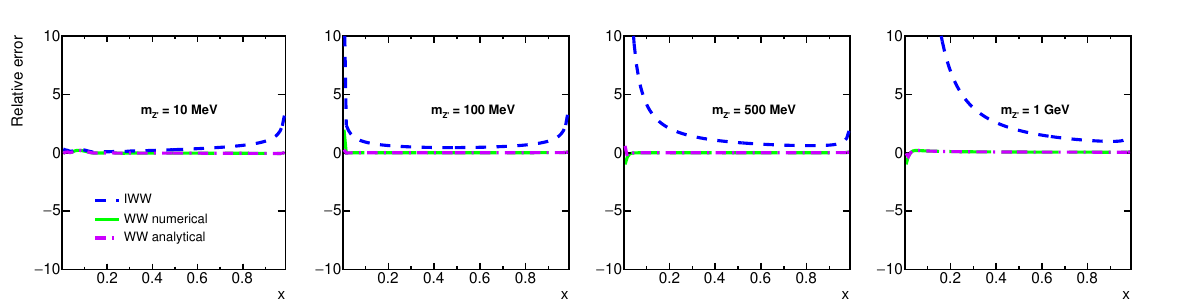}
    \caption{(Top) Single-differential cross-section $d\sigma_{2\rightarrow 3}^{X}/dx\rvert_\text{approx.}$ for the process
                   $\mu N\rightarrow\mu NZ'$ in both ETL, WW and IWW approximation. In the WW approximation, the results from the analytical 
                   expression of Eq. \eqref{eq:analytical-muon} and from the numerical integration of Eq. \eqref{eq:WWxtheta} are shown.
                   (Bottom) Relative error with respect to the ETL computation, defined as
                   $(\mathcal{O}_\text{approx.}-\mathcal{O}_\text{ETL})/\mathcal{O}_\text{ETL}$.}
    \label{fig:dsdx-muon}
\end{figure}

\begin{figure}[h]
    \centering
    \includegraphics[width=0.45\textwidth]{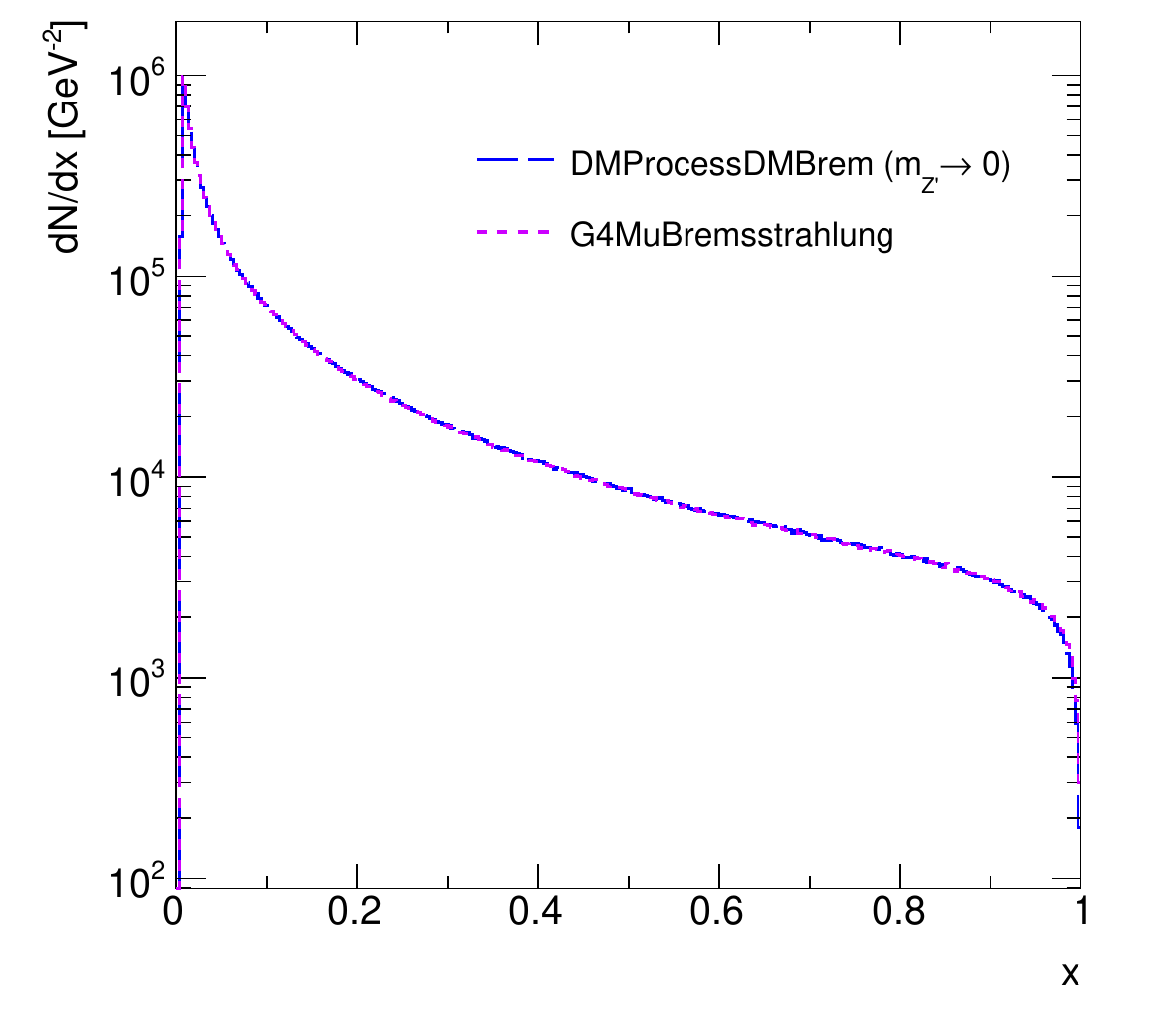}
    \caption{Comparison of the absolute yield for both SM bremsstrahlung, $\mu N\rightarrow\mu N\gamma$ (\texttt{G4MuBremsstrahlung}), and $Z'-$strahlung, $\mu N\rightarrow\mu NZ'$ (\texttt{DMProcessDMBrem}), in the limit $m_{Z'}\rightarrow 0$ as simulated with mono-energetic muons impinging on a thick block of lead. Both simulations are performed with Geant4, and light dark matter is generated through DMG4. Similar parameters and production cuts are chosen in both simulations.}
    \label{fig:G4brem-DMBrem}
\end{figure}

As introduced in Sec. \ref{sec:dark-matter}, the differential and total production cross sections for the reaction
$\mu N\rightarrow\mu NX$ can be computed within the WW approximation. Within this phase-space approximation, the $2\rightarrow 3$ process
depicted above is factorized through the use of the equivalent photon flux approximation \cite{Liu:2016mqv,Liu:2017htz} into
a $2\rightarrow 2$ process, provided the virtual photon flux
\begin{equation}
    \label{eq:photon-flux}
    \chi^\text{WW}=\int_{t_\text{min}}^{t_\text{max}}dt\ \frac{t-t_\text{min}}{t^2}F^2(t),
\end{equation}
with $t_\text{min}$ and $t_\text{max}$ being the minimum and maximum momentum transfer to the nucleus, and $F^2(t)$ the elastic
form factor associated with the nucleus (see e.g. \cite{Liu:2016mqv}). As such, the double-differential cross section
reads \cite{Kirpichnikov:2021jev}
\begin{equation}
\label{eq:WWxtheta}
\frac{d^2 \sigma^{X}_{2\to 3}}{d x d \cos\theta_{X}} \Bigr|_\text{WW} \simeq \frac{\alpha \chi^\text{WW}}{ \pi (1-x)} E_0^2 x \beta_{X}  \frac{d \sigma^{X}_{2\to 2} }{d (pk)} \Bigr|_{t=t_\text{min}},
\end{equation}
with $X=Z',S,P$, $x$ and $\theta_X$ being respectively the fractional energy of the $X$ boson and its emission angle, $\alpha$ the
fine structure constant, $\beta_X$ the corresponding Lorentz $\beta-$factor, $p$ and $k$ the four-momenta associated with the
initial-state muon and final-state $X$ boson, and $E_0$ the muon energy before the interaction. Together with the double differential
cross section of Eq. \eqref{eq:WWxtheta}, we implemented within the DarkMatter package the single-differential cross section for $x$,
obtained by performing a complicated analytical integration of Eq. \eqref{eq:WWxtheta}. This allowed us to obtain much better
run-time performance of the event sampling code. The total cross section then reads \cite{Sieber:2023nkq}
\begin{equation}
\label{eq:analytical-muon}
\sigma_{2\rightarrow 3}^{X}\Bigr|_\text{WW}=\int_{x_\text{min}}^{x_\text{max}}dx\ \bigg(\epsilon_X^2\alpha^3\sqrt{1-\frac{m_{X}^{2}}{E_0^2}}\frac{1-x}{x}\sum_{i=1}^{6}I_i^{X}(x,\tilde{u})\bigg\rvert_{\tilde{u}=\tilde{u}_\text{min}}^{\tilde{u}=\tilde{u}_\text{max}}\bigg),
\end{equation}
where the six special functions $I_i$, $i=1,2,...6$ and the definition of the variable $\tilde{u}_\text{min}$ and $\tilde{u}_\text{max}$
can be found in \cite{Sieber:2023nkq}. This provides one with a more precise expression than within the IWW approximation, where the dependence on 
$(x,\ \theta_X)$ is neglected in Eq. \eqref{eq:photon-flux}. As a result, we don't need K-factors and tables for muon-beam-related simulations.
For an appreciable estimate of the precision of the analytical results of the WW approximation
integration, a comparison of the single-differential cross section $d\sigma_{2\rightarrow 3}^{X}/dx\rvert_\text{WW}$ with both ETL
and IWW results is shown in Fig. \ref{fig:dsdx-muon} for different $m_{X}$ in the $X=Z'$ scenario. The relative error between
the computations at ETL and within the WW phase-space approximation is $\leq2\%$, while both the WW numerical and analytical methods
agree well within $<1\%$. The reader is referred to \cite{Kirpichnikov:2021jev} and \cite{Sieber:2023nkq} for a more in-depth discussion
of the errors.\\ \indent
As a final test, the implementation of the $Z'$ vector boson emission process in the mass limit $m_{Z'}\rightarrow 0$ is compared
to the Geant4 treatment of SM muon bremsstrahlung, $\mu N\rightarrow\mu N\gamma$. The comparison is driven against the absolute yields of both
processes. As such, those are extracted from a minimal Geant4 simulation of muons impinging on a thick block of lead (Pb).
The parameters and production cuts of the \texttt{G4MuBremsstrahlungModel} are chosen to be comparable with the one of \texttt{DarkZ},
namely similar values of $t_\text{min}$ and $t_\text{max}$, and equal parameters $x_\text{min}$ and $x_\text{max}$
(respectively $v_\text{cut}$ and $v_\text{max}$ within Geant4, see \cite{Bogdanov:2006kr}), as well as a 1 GeV production cut
(emission threshold). This later choice implies that the cross sections do not depend on the $Z'$ mass for values below 0.1 MeV.
The results for mono-energetic 160 GeV muons with $g_{Z'}=\epsilon_{Z'}e=1$ are shown in Fig. \ref{fig:G4brem-DMBrem}. Within
the statistical uncertainties, it is found that the absolute yields agree within $1\%$.

\subsection{Final-state muon kinematics}

\begin{figure}[h]
    \centering
    \includegraphics[width=0.45\textwidth]{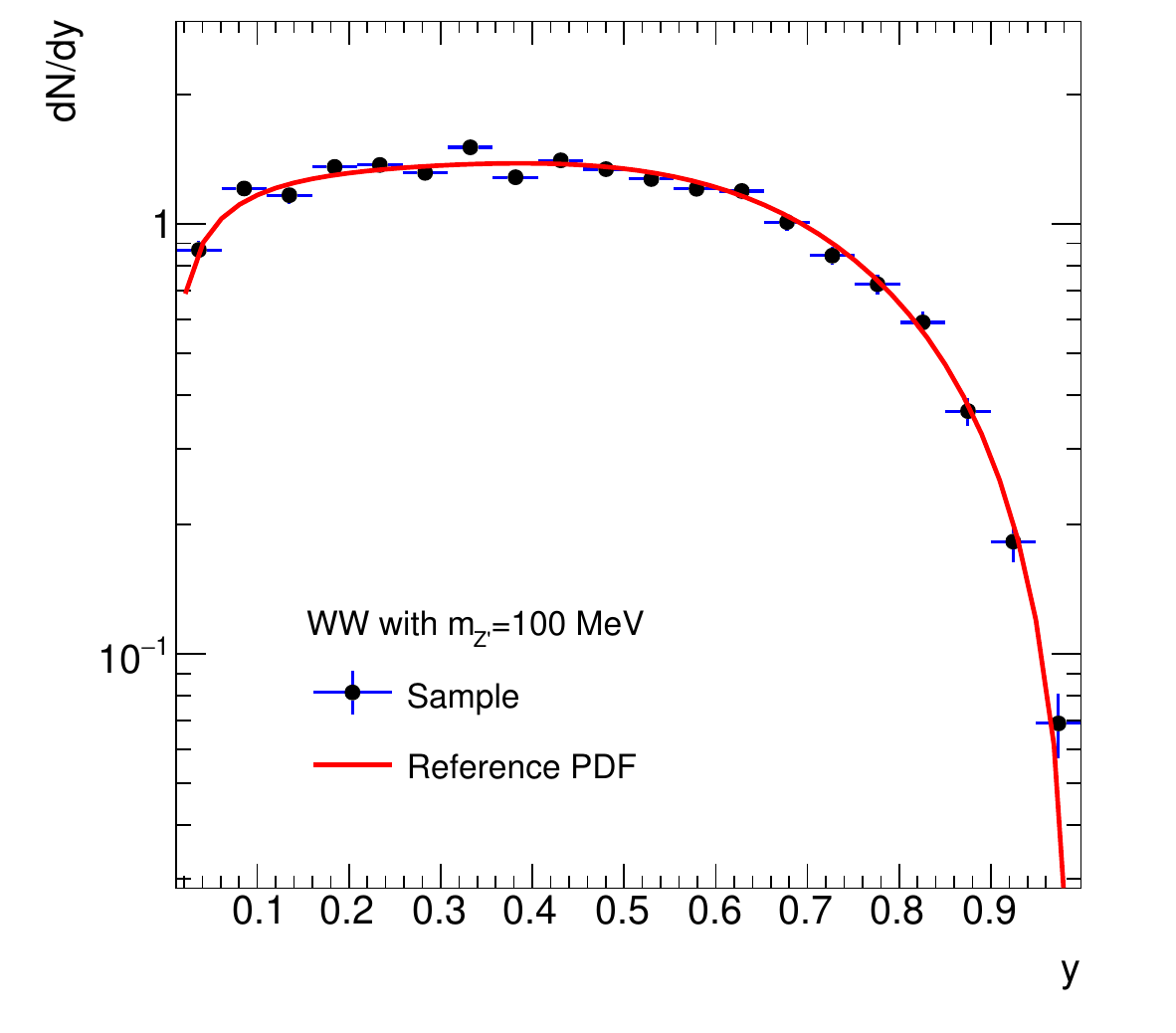}
    \includegraphics[width=0.45\textwidth]{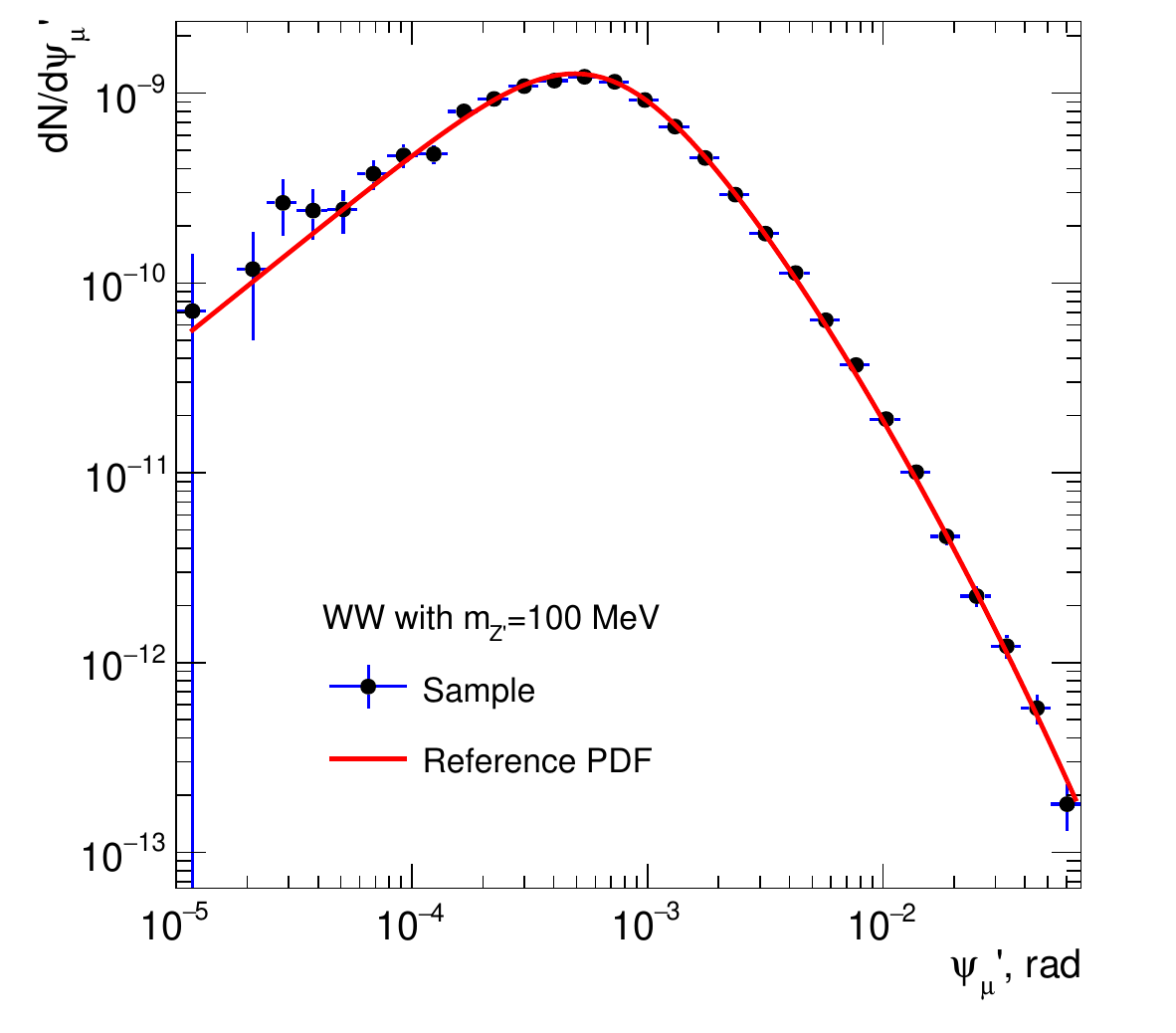}
    \caption{Comparison between the target partial distribution function and the sampled events from the DarkMatter function \texttt{SimulateEmissionByMuon2} for final-state muons within a DMG4-based simulation of the process $\mu N\rightarrow\mu NZ'$. (Left) The fractional muon energy distribution, $y$. (Right) The final-state muon emission angle distribution, $\psi_\mu^\prime$.}
    \label{fig:muon-sampling}
\end{figure}

As compared to the experiments in electron/positron beams, for the muon beam the sampling of the recoil muon kinematic variables
is much more important. This is due to the fact that the missing energy in this case is determined by measuring the recoil muon
momentum by tracker. Similarly to Eq. \eqref{eq:WWxtheta}, for the recoil muon variables we have
\begin{equation}
\label{eq:WWypsi}
\frac{d^2 \sigma^{X}_{2\to 3}}{d y d \cos\psi_{\mu}^\prime} \Bigr|_\text{WW} \simeq \frac{\alpha \chi^\text{WW}}{ \pi (1-y)} E_0^2 y \beta_{\mu}^\prime  \frac{d \sigma^{X}_{2\to 2} }{d (pp')} \Bigr|_{t=t_\text{min}},
\end{equation}
where $y$ is the final-state muon fractional energy, $\psi_\mu^\prime$ is its emission angle, $\beta_\mu^\prime$ is its Lorentz
$\beta-$factor and $p^\prime$ is its four-momentum. The corresponding double differential cross section is implemented in the package
DarkMatter. For illustrative purposes, the results of events sampling for both the $y$
and $\psi_\mu^\prime$ variables are shown Fig. \ref{fig:muon-sampling}, as extracted from a DMG4-based simulation of light
$Z'$ production with fixed mass $m_{Z'}=100$ MeV. The validity of the implementation is inferred through a Kolmogorov-Smirnov
(KS) test \cite{Kolmogorov:1933sde,Smirnov:1948tfe} for the goodness of the fit, comparing both the target partial distribution
function (PDF) and the corresponding sampled histogram, and indicating for the test mass range $m_{Z'}=10-1000$ MeV that
the identity null hypothesis is accepted at a 0.05 significance level.

\section{Developments in the annihilation processes}

In this section, we describe the main DMG4 developments and improvements concerning light dark matter production via resonant
$e^+e^-$ annihilation. We refer the reader to our previous work~\cite{Bondi:2021nfp} for a comprehensive discussion regarding
the general implementation of this process in the package.

\subsection{Extension to different models}

For the four mediator cases already implemented in the DMG4 package, $e^+e^- \rightarrow X \rightarrow \chi \chi$, where $X$
can be a vector, axial-vector, scalar, and pseudo-scalar particle, we introduced the possibility for the final state DM particles
$\chi$ to be (complex) scalars. Specifically, we considered the following Lagrangians, where $\Phi$ is the LDM complex scalar
field, $\psi$ is a SM charged field. $S$, $P$, $A_\mu$, $V_\mu$ are the mediators field, while $\epsilon_V, \epsilon_A,
\epsilon_S, \epsilon_P$ are the mixing (or coupling) parameters. Finally, $m_V,m_A,m_S,m_P$ are the masses of the mediators.
\\
Vector case: 
\begin{equation}
 \mathcal{L} \supset \mathcal{L}_{SM} -\frac{1}{4} V_{\mu\nu}^2 +\frac{1}{2} m_V^2 V_\mu^2
  - \sum_{\psi} e \epsilon_V V_\mu  \bar{\psi} \gamma^\mu \psi
  + g^D_V  V_\mu (\Phi^* \partial^\mu \Phi - (\partial^\mu \Phi)\Phi^*)
  + \frac{1}{2} (\partial_{\mu}\Phi)^2 - \frac{1}{2}m^2_\Phi \Phi^*\Phi 
 \label{VecLagr1}
\end{equation}
Axial vector case:
\begin{equation}
  \mathcal{L} \supset \mathcal{L}_{SM} -\frac{1}{4} A_{\mu\nu}^2 +\frac{1}{2} m_A^2 A_\mu^2
  - \sum_{\psi} e \epsilon_A A_\mu  \bar{\psi} \gamma_5 \gamma^\mu \psi
  + g^D_A  A_\mu (\Phi^* \partial^\mu \Phi - (\partial^\mu \Phi)\Phi^*)
  + \frac{1}{2} (\partial_{\mu}\Phi)^2 - \frac{1}{2}m^2_\Phi \Phi^*\Phi
 \label{AxialVecLagr1}
\end{equation}
Scalar case:
\begin{equation}
  \mathcal{L} \supset \mathcal{L}_{SM} +\frac{1}{2} (\partial_{\mu}S)^2 - \frac{1}{2}m^2_S S^2
  - \sum_{\psi} e \epsilon_S S\bar{\psi}{\psi}
  + g^D_S m_S S \Phi^* \Phi
  + \frac{1}{2} (\partial_{\mu}\Phi)^2 - \frac{1}{2}m^2_\Phi \Phi^*\Phi
 \label{ScalLagr1}
\end{equation}
Pseudo-scalar case:
\begin{equation}
  \mathcal{L} \supset \mathcal{L}_{SM} +\frac{1}{2} (\partial_{\mu}P)^2 - \frac{1}{2}m^2_P P^2
  - \sum_{\psi} e \epsilon_P P\bar{\psi}\gamma^5{\psi}
  + g^D_P m_P P \Phi^* \Phi
  + \frac{1}{2} (\partial_{\mu}\Phi)^2 - \frac{1}{2}m^2_\Phi \Phi^*\Phi \; \;.
 \label{PseudoscalLagr1}
\end{equation}

The corresponding cross sections for the $e^+e^- \rightarrow X \rightarrow \Phi \Phi$ processes implemented in the code reads
\begin{equation}
\sigma_{\ee}=\frac{4\pi \alpha_{EM} \alpha_D \varepsilon^2}{\sqrt{s}}q\frac{\mathcal{K}}{(s-m_X^2)^2+\Gamma_X^2m_X^2} \; \;
\end{equation}
where $s$ is the invariant mass of the $\ee$ system, $m_X$ the mediator mass, $q=\frac{\sqrt{s}}{2}\sqrt{1-\frac{4m_\Phi^2}{s}}$,
$\Gamma_X$ is the intermediate DM particle decay width to dark particles $\Phi$, $\alpha_{EM}$ is the electromagnetic fine
structure constant, and $\alpha_D\equiv \frac{\left(g^D_X\right)^2}{4\pi}$ is the $\Phi$ coupling squared to the dark particles $\chi$.
Finally, $\mathcal{K}$ is a kinematic factor that reads, respectively, $2/3 q^2$ for the vector and axial-vector mediator,
and $m^2_X/4$ for the scalar and pseudo-scalar cases. Finally, the $X$ decay widths read:
\begin{eqnarray}
\Gamma_{V\rightarrow \Phi^* \Phi} & =& \frac{\alpha_D}{12} m_{V}\bigl(1-4r^2\bigr)^\frac{3}{2}
\\
\Gamma_{A\rightarrow \Phi^* \Phi} &= &\frac{\alpha_D}{12} m_{A}\bigl(1-4r^2\bigr)^\frac{3}{2} 
\\
\Gamma_{S\rightarrow \Phi^* \Phi} &= &\frac{\alpha_{D}}{4} m_{S}\bigl(1-4r^2\bigr)^{1/2}
\\
\Gamma_{P\rightarrow \Phi^* \Phi} &= &\frac{\alpha_{D}}{4} m_{S}\bigl(1-4r^2\bigr)^{1/2},
\end{eqnarray}
where $r=m_\Phi/m_X$.

For the vector-mediator model, we also considered an inelastic (i.e., non-diagonal) model foreseeing two fermionic LDM states
$\chi_2$ and $\chi_1$ with different masses $m_2$ and $m_1$, with $\Delta = m_2 - m_1 > 0$~\cite{NA64:2021acr}. In this model,
at tree level only the $A^\prime - \chi_2 - \chi_1$ vertex exists. The corresponding Lagrangian reads:

\begin{equation}
  \mathcal{L} \supset \mathcal{L}_{SM} -\frac{1}{4}V_{\mu\nu}V^{\mu\nu}+\frac{1}{2}m^2_VV^2
  - \sum_{\psi} e \varepsilon_V V^\mu \bar{\psi}\gamma_\mu{\psi}
  + \sum_{i}\bar{\chi}_i(\slashed{\partial}-m_i)\chi_i
  + g_D A^\mu (\bar{\chi}_2\gamma_\mu\chi_1 + \bar{\chi_1}\gamma_\mu \chi_2)
 \label{NonDiagLagr1}
\end{equation}

The corresponding annihilation cross section reads:
\begin{equation}
\sigma_{\ee} = \frac{8\pi \varepsilon_V^2 \alpha_D\alpha_{EM}}{\sqrt{s}}\frac{q}{(s-m^2_V)^2+\Gamma^2m^2_V}
(m_1m_2+E_1E_2-\bar{p}^2/3)\;\;,
\end{equation}
where $q$ is the $\chi_2$ and $\chi_1$ three-momentum in the $V$ rest frame, while $E_2$ and $E_1$ are the final state particles
total energies in the same frame. The decay width $\Gamma$ for the $V\rightarrow \chi_2 \chi_1$ channel is:
\begin{equation}
  \Gamma_{V\rightarrow \chi_{2}\chi_{1}} = \frac{\alpha_D}{3m^2_V} 2q
  \left(
  3m_1m_2+m^2_V-\frac{m^2_1+m^2_2}{2}-\frac{(m^2_2-m^2_1)^2}{2m^2_V}
  \right)
\end{equation}

In this DMG4 software version, we also included the possibility to simulate the
$e^+e^- \rightarrow Z^\prime \rightarrow \bar{\nu}\nu$ process, where the intermediate $Z^\prime$ particle is the massive force
mediatior associated with a new SM gauge symmetry involving the neutrino sector. Specifically, we considered the two cases
of a $B-L$ gauge symmetry and a $L_\mu-L_\tau$ one \cite{NA64:2022rme,NA64:2022yly}. While the first case resembles that
of a vector dark photon by making the substitution $\varepsilon e \leftrightarrow g$, where $g_{\Zpr}$ is the gauge coupling
constant, for the second scenario at tree level the $Z^\prime$ couples only to second and third generation leptons.
Nevertheless, thanks to the photon-$Z^\prime$ coupling introduced at next-to-leading order by loops involving muon and tau
leptons, an effecting $Z^\prime-e^+-e^-$ vertex appears, with coupling $e \Pi(q^2)$, where $q^2$ is the $Z^\prime$
four-momentum squared. The function $\Pi$ reads:
\begin{equation}
\label{eq:pi}
    \Pi(q^2)=\frac{e\,g_\Zpr}{2\pi^2}\int_0^1dx \, x (1-x) \ln\frac{m^2_\tau-x(1-x)q^2}{m^2_\mu-x(1-x)q^2}\;\;.
\end{equation}
In the code, we were able to introduce the full analytical expression for the $\Pi$ function, allowing to optimize the corresponding
computation time. This is given by the formula:
\begin{equation}
\begin{split}
\frac{2 \pi^{2}}{e g_{Z^{\prime}}} \Pi\left(q^{2}\right)=
\frac{1}{3}\left[\frac{1}{2} \log \left(\frac{r_{\tau}}{r_{\mu}}\right)+2\left(r_{\mu}- r_{\tau}\right)+ \right. \\
 -\left(1+2 r_{\mu}\right) \sqrt{1-4 r_{\mu}} \operatorname{coth}^{-1}\left(\sqrt{1-4 r_{\mu}}\right)+\\
\left.+\left(1+2 r_{\tau}\right) \sqrt{1-4 r_{\tau}} \operatorname{coth}^{-1}\left(\sqrt{1-4 r_{\tau}}\right)\right],\quad
\end{split}
\end{equation}
with $r_{\mu}=m_{\mu}^{2} / q^{2}$ and $r_{\tau}=m_{\tau}^{2} / q^{2}$.

\subsection{Narrow-width resonance production}

The introduction of all new DMG4 processes in Geant4 is based on the \texttt{G4VDiscreteProcess} class - each process is described
by a new class inheriting from the latter. In particular, each class has to provide a concrete implementation of the two pure
virtual methods \texttt{GetMeanFreePath} and \texttt{PostStepDoIt}; in particular, the first method is responsible of computing
the positron mean free path for the annihilation process at the beginning of each Geant4 step,
\textit{when the step-length is not yet determined}, from the knowledge of the corresponding \texttt{G4Track}. It follows that,
if the cross section for the annihilation process is computed solely from the value of the positron energy at the beginning
of the step $E_i$, without accounting for the full cross-section energy dependence, the obtained result would under-estimate
(over-estimate) the average value along the step if $E_i>E_\text{MAX}$ ($E_i<E_\text{MAX}$), where $E_\text{MAX}$ is the positron resonant
energy in the laboratory frame. While in the case of a broad resonance this effect is not critical, if the typical energy
loss across a Geant4 step is significantly larger than the mediator width $\Gamma$, it could induce a significant distortion
of the simulation results in case of a narrow resonance - this is the case, for example, of the $Z^\prime$ model described
previously in the text. Two ad-hoc modifications were introduced in the code to account for this effect.

\subsubsection{Energy dependence of the cross section}\label{sec:ene_dep}

As discussed previously, the \texttt{G4VDiscreteProcess}-based implementation requires to compute the cross section for the
annihilation process at the beginning of each Geant4 step, before the latter is determined. To account for the energy dependency
of the cross section, we followed the strategy implemented in Geant4 for other electromagnetic-like processes and described in
the manual, based on the sole assumption that the cross section $\sigma(E)$ has only one maximum $\sigma_\text{MAX}$, for $E=E_\text{MAX}$.
Specifically, our implementation assumes that, during the step \textit{still to be computed}, the cross section $\sigma(E)$ will be
lower than a certain value $\sigma_m$. The mean free path for the process is then computed via $\sigma_m$. If this process is selected
by Geant4 as that taking place and limiting the next step length, the final state sampling, implemented via the \texttt{PostStepDoIt}
method, is invoked with reduced probability $p_{red}=\sigma(E_f)/\sigma_m$, where $E_{f}$ is the positron energy at the end of the step;
alternatively, nothing happens and the positron tracking continues. In the following we call this procedure ``Maximum-Rejection''.
We underline that it is not obvious that in this procedure $p_{red}$ should be computed considering the cross section at the end of the
step and not the average value along the step\footnote{Incidentally, we note that in the Geant4 physics reference manual for release 11.2,
rev 8.0, in Sec. 7.4 \textit{``Correcting the Cross Section for Energy Variation''} the prescription to use of $\sigma(E_f)$ in the
computation of $p_{red}$ is supported by a reference to ~\cite{Ivanchenko:1991me}. However, in that work no justification
of this method is provided.}. To empirically verify this, we made a dedicated toy-MC study, presented in Appendix A.

\begin{figure}
 \centering
  \includegraphics[width=.6\textwidth]{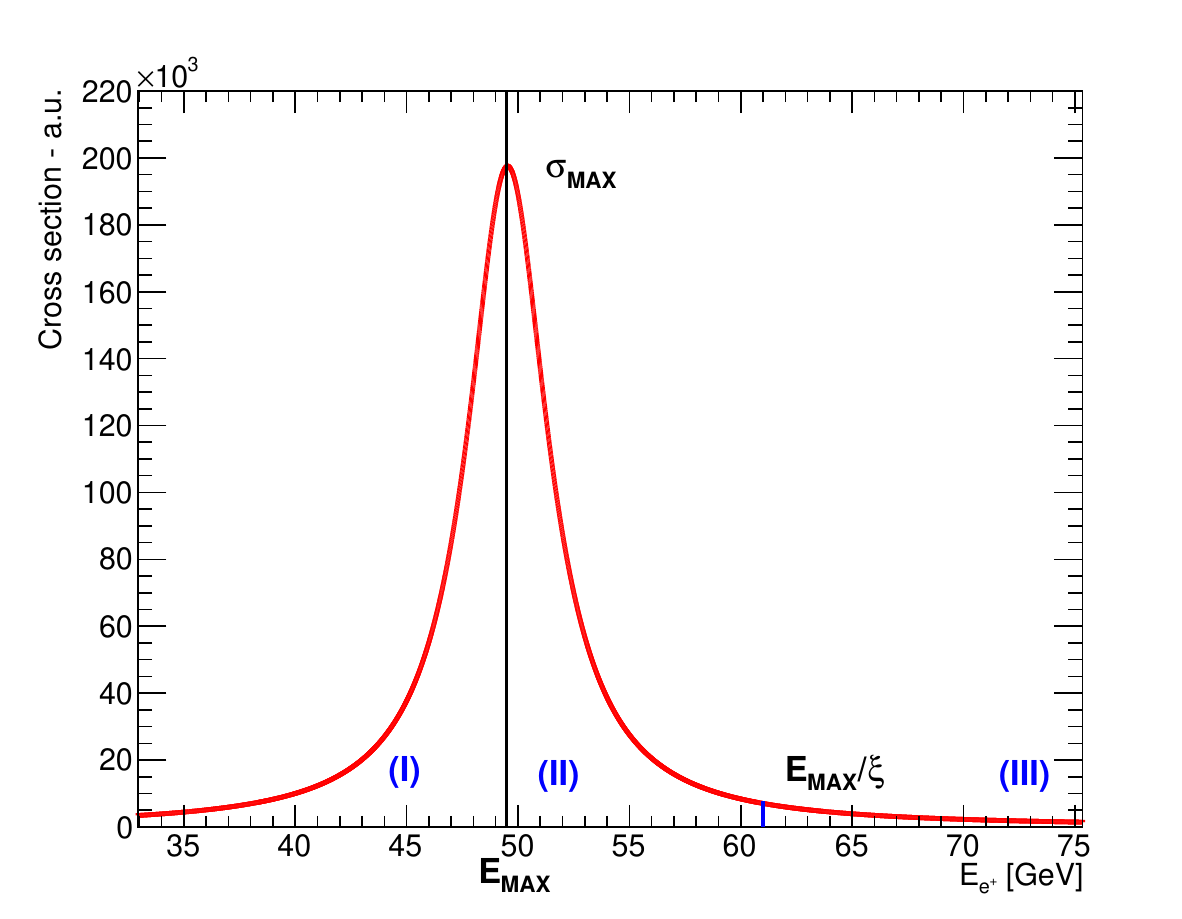}
  \caption{\label{fig:annihilation0}Cross section $\sigma(E_+)$ for the annihilation process $e^+e^-\rightarrow X \rightarrow \chi \chi$
for a narrow-width mediator. The figure highlights the three regions (I), (II), and (III) in which the energy interval is divided
into to handle the energy dependence of the cross section.}
\end{figure}

The value $\sigma_{m}$ is selected as follows, dividing the positron energy range in three dinstinct intervals, also shown
in Fig.~\ref{fig:annihilation0}. The first case corresponds to the condition $E_i<E_{MAX}$: in this case, the positron energy
is already lower than the cross section maximum, and $\sigma(E)$ will further decrease along the positron trajectory;
therefore, $\sigma_m=\sigma(E_i)$. In the second case, instead, $E_{MAX}<E_i<E_{MAX}/\xi$, where $0<\xi<1$ is a dimensionless
parameter to be appropriately tuned; this case corresponds to a positron with initial energy close to that of the resonance,
and assumes $\sigma_m=\sigma_{MAX}$. Finally, if $E_i>E_{MAX}/\xi$, then $\sigma_m=\sigma(\xi E_i)$.

The parameter $\xi$ controls the transition between case (II) and case (III), and is connected to the typical positron energy loss
across a step. If the typical value of the \textit{continuous} energy loss across a step is $\Delta E$, then the value of $\xi$
can be set from the requirement that, for a positron initial energy $E_i$ in the interval $[E_R,E_i-n \Delta_E]$, where $n>1$
is a safety factor, case (II) should be applied, since it is possible that, in the next step, the particle will loose enough
energy to cross the resonant energy. All values of $\xi$ in the interval $[\frac{E_R}{E_R+n(\Delta E)_{MAX}},1]$ satisfy
this requirement. In the code, the lower limit of the interval is used.

\subsubsection{Dynamical step limitation}

The approach presented in the previous section only considers \textit{continuous} energy losses for the positron, but not
the \textit{discrete} ones associated to bremsstrahlung emission, that may significantly alter the positron energy.
Also, this approach is expected to work well for geometries with many geometrical boundaries, forcing many positron steps,
but can be inaccurate for large homogeneous volumes. To partially compensate for this, we implemented a step limitation
for positrons. The intrinsic scale of the positron step limitation $\delta_0$ for narrow resonance production
is given by the ratio of the modified resonance width $\Gamma^*=\Gamma M_X/m_e$, where the factor $M_X/m_e$ is a kinematic
correction induced by the Lorentz transformation from the center-of-mass frame to the laboratory one, and the positron energy
loss per unit length $-\frac{dE}{dx}$, $\delta_0=-\Gamma^*/\frac{dE}{dx}$.

In order to optimize the computation time, we use a dynamical step limitation for positrons, based on the relative cross
section variation with respect to the energy: the more the cross section is expected to vary for a given positron energy,
the smaller the maximum step allowed for the positron. The resonant annihilation cross section can indeed be written as
$\sigma = f(E_+) \frac{1}{(E-E_{MAX})^2+(\Gamma^*)^2}$, where $f$ is a smooth-varying function of the positron energy.
It follows that the relative cross section variation with respect to the energy reads:
\begin{equation}
V(E)=\frac{1}{\sigma}|\frac{d\sigma}{dE}| \simeq \frac{2|E-E_R|}{(E-E_{R})^2+(\Gamma^*)^2}\;\;.
\end{equation}
This function has two maxima at $E=E_{R}\pm \Gamma^*$, where $V=1/\Gamma^*$. To avoid numerical instabilities in the interval
between these two values, associated with the zero at $E=E_R$, we slightly modified it to be $V^\prime=V$ if $|E-E_{R}|>\Gamma^*$,
and $V^\prime=1/\Gamma^*$ otherwise. In the code, we thus define the positron step limit as
$\delta_{MAX}=\delta_0  \cdot 1/V^\prime \cdot (1/\Gamma^*)$.

\subsection{Atomic effects in $e^+e^-$ annihilation}

In the previous DMG4 version, all $e^+e^-$ annihilation processes were handled by considering the atomic electron to be at rest in
the laboratory frame. Atomic motions reflect to a broadening of the annihilation cross section $\sigma(E_+)$, since a given positron
energy can reflect to different values of the Mandelstam variable $s$ for the electron-positron system, defined by the equation:
\begin{equation}
s=2m_e^2+2E_+(E_--zP_-)\;\;,
\end{equation}
where $(E_-)$ ($P_-$) is the atomic electron total energy (three-momentum), and $z$ is the cosine of the angle between the $e^+$
and $e^-$ momenta. From this, it follows that a positron whose energy won't match the resonance mass if annihilating with an at-rest
electron can do so for an orbiting one and vice-versa, resulting to a broad range $\Delta_E$ of energies that can contribute to the process.
To illustrate this effect, we consider the case of a $m_X=225$ MeV mediator, and an electron kinetic energy of about 10 keV
(typical value for lead): the positron energy in the laboratory frame resulting to $s=m_X^2$ spans a range from 40.7 GeV to 60.3 GeV
(for comparison, the at-rest value is 48.6 GeV). Similarly, for $m_X=17.7$ MeV, considering an electron kinetic energy of about 290 eV
(typical value for carbon), the resonant positron energy in the laboratory frame spans from 273 MeV to 291 MeV. From this consideration,
it follows that the at-rest approximation is justified for large-width resonances, when $\Gamma_X \frac{m}{m_e}\gtrsim \Delta_E$,
while it can't be applied in the narrow width scenario.

In this DMG4 software version, we implemented a full calculation involving atomic motions as follows. For the simpler case
of massless final state particles, such as in the reaction $e^+e^- \rightarrow Z^\prime \rightarrow \nu \bar{\nu}$,
the general expression for the cross section can be written as $\sigma = A(E_+) \frac{1}{(s-m^2_X)^2+m^2_X\Gamma^2}$,
where $A$ is a smooth-varying function of the impinging positron energy, and $m_X$, $\Gamma$ are, respectively,
the mediator mass and total width. By substituting the previous expression for $s$, this translates to:
\begin{equation}
\sigma = f(E_+) \frac{1}{(2m_e^2+2E_+(E_--zP_-)-m^2_X)^2+m^2_X\Gamma^2} \; \;.
\end{equation}
The code computes, for each interaction, the average value of this expression, by considering an isotropic electron motion,
i.e. a flat PDF for $z$ in the $[-1,1]$ interval:
\begin{align}\label{eq:barsigma}
  \bar{\sigma} &= \int_{-1}^{1}dz \,p(z)\, \frac{A(E_+)}{(2m_e^2+2E_+(E_--zP_-)-m^2_X)^2+m^2_X\Gamma^2} =
  \\ &=\frac{A(E_+)}{4m_X\Gamma}\left[
    \arctan{\frac{m^2_X+2P_+P_--2m^2_e-2E_+E_-}{m_X\Gamma}}
    -
    \arctan{\frac{m^2_X-2P_+P_--2m^2_e-2E_+E_-}{m_X\Gamma}}
    \right]\;\;,
\end{align}
where $E_+$ and $E_-$ ($P_+$ and $P_-$) are the positron and the electron total energies (momenta) in the laboratory frame,
and $p(z)=\frac{1}{2}$ within the integration interval.

\begin{figure}[t]
  \centering
  \includegraphics[width=.6\textwidth]{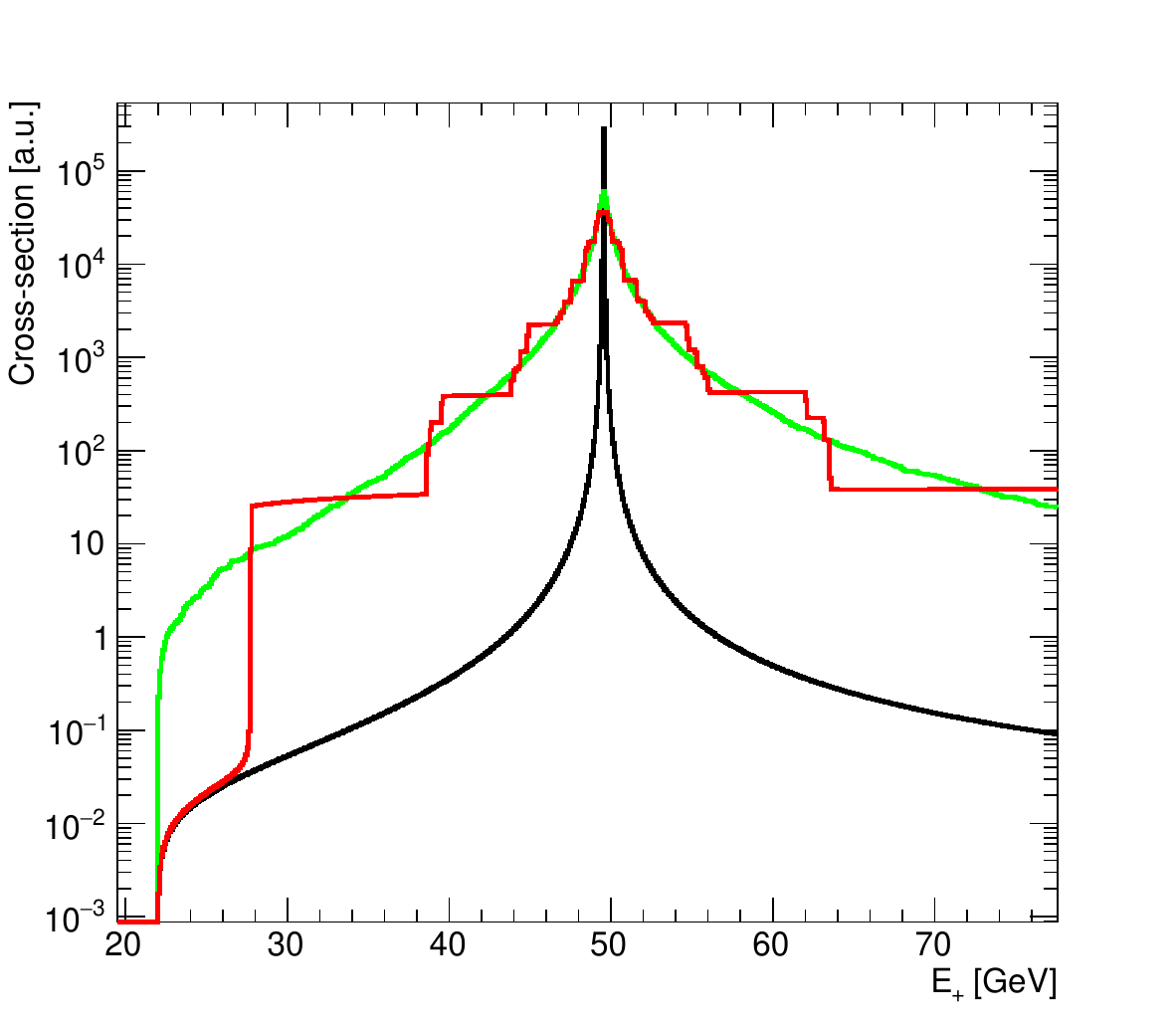}
  \caption{\label{fig:annihilation1}Cross section $\sigma(E_+)$ for the annihilation process $e^+e^-\rightarrow Z' \rightarrow \nu \bar{\nu}$
 for a 225 MeV $Z^\prime$ with $\Gamma=7$ keV, decaying into neutrinos. The black curve is the result obtained without including atomic
 effects, while the red (green) curve are the results obtained from the simplified (realistic) atomic model currently implemented in DMG4.}
\end{figure}

Concerning the atomic electron energies, the code currently implements two possible models. The first, simplified model assumes,
for each atomic shell, a fixed kinetic energy equal to the binding energy magnitude, $T_-=|B|$, and computes the average cross
section by summing over the atomic shells, considering for each a weight proportional to the occupation number. The shells
atomic energies for the target material are obtained directly from Geant4, using the \texttt{G4Element::GetAtomicShell} method.
Similarly, the occupation numbers are obtained from the \texttt{G4Element::GetNbOfShellElectrons} method. In the second,
more realistic model, used by default, an exponential PDF distribution~\footnote{We incidentally note that this model
is the same adopted by the Geant4 software to handle atomic motion effects in Compton scattering.} for $T_-$ is considered,
$p(T_-)=\frac{1}{|B|}\exp(-T_-/|B|)$. The contribution for each shell is separately computed, by integrating the expression
for $\bar{\sigma}$ over $p(T_-)$, and summing the contributions with weights each equal to the shell occupation number.
The code computes the integral over $p(T_-)$ numerically through the Monte Carlo method. To speed-up the computation time,
since the expression for $p(T_-)$ is independent from the positron kinematics, a set of random electron kinetic energies
${T_-}$ is pre-generated for each atomic shell at initialization time. The effect of the inclusion of atomic motion
description in the software is illustrated in Fig.~\ref{fig:annihilation1}, showing the annihilation cross section
as a function of the positron energy for a 225 MeV resonance, having a width of 7 keV. The black curve is the result
obtained without considering atomic motions, while the red (green) curve are the results obtained from the first and second
model described before. In the first case, the discreteness of the atomic shells kinetic energies is clearly visible.

\begin{figure}[t]
  \centering
  \includegraphics[width=.6\textwidth]{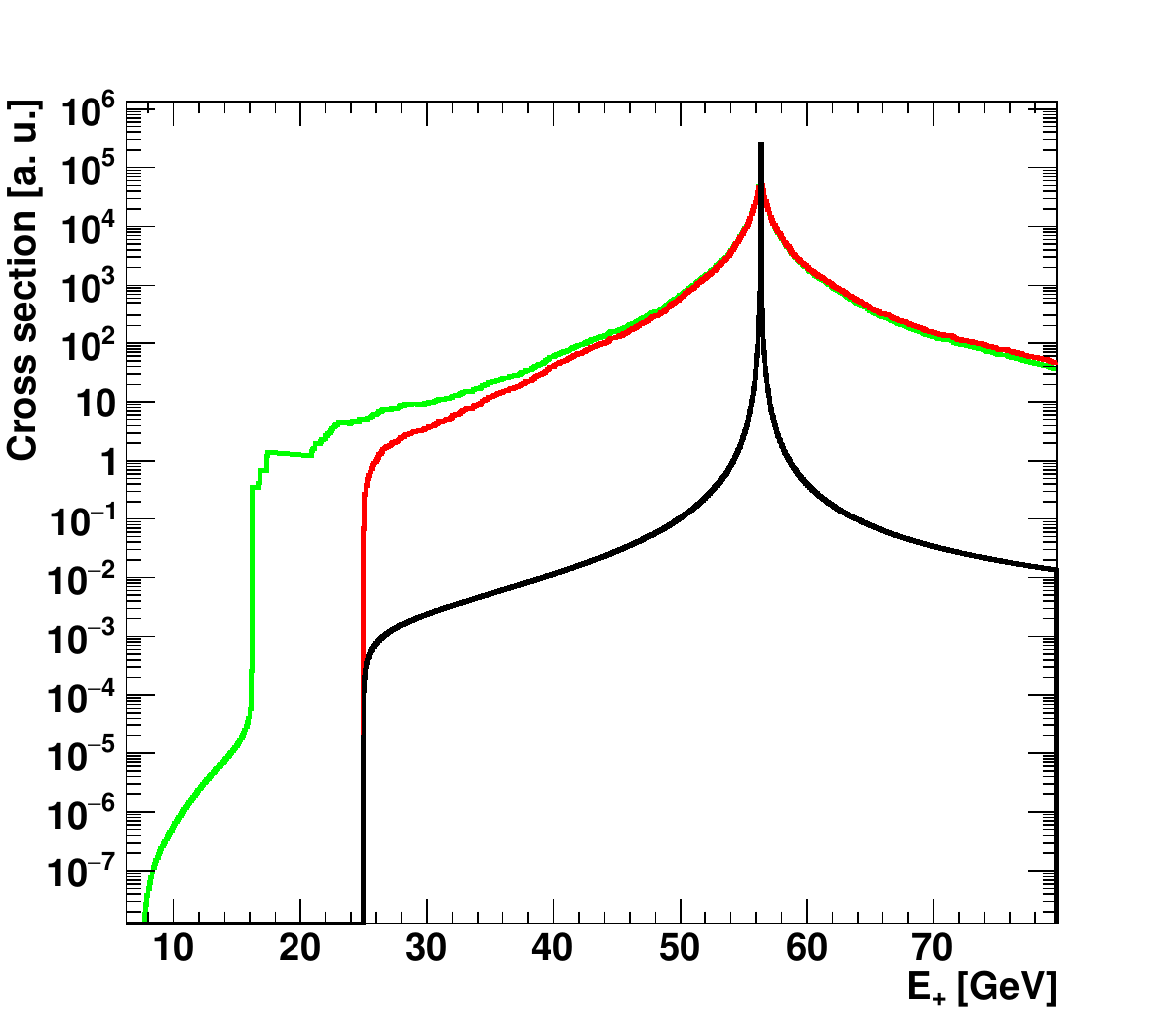}
  \caption{\label{fig:annihilation2}Cross section $\sigma(E_+)$ for the annihilation process $e^+e^-\rightarrow X \rightarrow \chi \chi$ for a 
240 MeV mediator with $\Gamma\simeq1$ keV, decaying into 80 MeV final state particles, after including atomic effects.
The black curve is the result obtained without considering atomic effects, the red curve is the result obtained ignoring the $q$ dependency
of the cross section, and finally the green curve is the full result accounting for the kinematic threshold. Electron atomic energies
were simulated using the realistic model previously described.}
\end{figure}

This approach is slightly modified for the more general case involving equally massive particles in the final state, such as
in the LDM production process $e^+e^-\rightarrow X \rightarrow \chi \chi$, since in this case the function $A(E_+)$ is proportional
to the final-state momentum $q$, that intrinsically contains a kinematic threshold\footnote{The extension of the code to handle
final state with particles having different masses is foreseen for the next DMG4 release.}. Starting from the previous
for $\bar{\sigma}$ in Eq.~\ref{eq:barsigma}, and writing explicitly the $q$ factor, the expression for $\bar{\sigma}$ is:
\begin{equation}
 \bar{\sigma} = \int_{z_{min}}^{z_{max}}dz \,p(z)\, \frac{q B(E_+)}{(2m_e^2+2E_+(E_--zP_-)-m^2_X)^2+m^2_X\Gamma^2} \; \;,
\end{equation}
where $B(E_+)\equiv A(E_+)/q$. The integration range can more easily be determined by changing the integration variable from $z$
to $s$ for fixed $E_+$, and explicitly setting $p(z)=\frac{1}{2}$:
\begin{equation}
 \bar{\sigma} = \frac{1}{4E_+P_-}\int_{s(z=z_{max})}^{s(z=z_{min})} ds \frac{B(s)q(s)}{(s-M^2)^2+M^2\Gamma^2} \; \; ,
\end{equation}
and considering the threshold requirement $s>s_{min}=4m^2_\chi$. From the identity $s=2m_e^2+2E_+(E_--zP_-)$, it follows
that $z<z_{max}=\frac{2m^2_e+2E_+E_- - s_{min}}{2E_+P_-}$. Note that if $z_{max}<-1$, then the positron energy is incompatible
with $s_{min}$ for all angles, and thus $\bar{\sigma}=0$. If, instead, $z_{max}>1$, the integral is still performed only
in the physical region up to $z_{max}=+1$. The result of the integral, assuming a constant value for $B$
and defining $x=\sqrt{s-4m^2_\chi}$ is:

\begin{equation}
\bar{\sigma}=\frac{B}{8E_+P_-} \cdot \left[P(x(s(z=-1)))-P(x(s(z=z_{max})))\right] \; \;, 
\end{equation}
where $P(x)$ is given by:
\begin{align}
P(x)=&
\left[
\frac{1}{2\sqrt{2(\eta^2+\Delta^2)}}
\log{\frac{x^2-\sqrt{2(\eta^2+\Delta^2)}x+\eta^2}{x^2+\sqrt{2(\eta^2+\Delta^2)}x+\eta^2}}\right.
+
\\
&
+\frac{1}{2\sqrt{\eta^2/2-\Delta^2/2}}
\left.
\left(
\arctan{\frac{2x-\sqrt{2(\eta^2+\Delta^2)}}{\sqrt{2\eta^2-2\Delta^2}}}
+
\arctan{\frac{2x+\sqrt{2(\eta^2+\Delta^2)}}{\sqrt{2\eta^2-2\Delta^2}}}
\right)
\right] \; \;,
\end{align}
where $\Delta^2=m_X^2-4m^2_\chi$ and $\eta^4=\Delta^4+\Gamma^2m^2_X$. Finally, to handle the residual $s-$ (and thus $z-$) dependency
in $B$ for fixed values of $E_+$ and $E_-$, this term is computed event-by-event via Monte Carlo integration, taking 20 uniformly-spaced $z-$
values in the allowed interval and averaging over the corresponding $B$ values. As an example, Fig.~\ref{fig:annihilation2} shows
the obtained cross section, considering the more realistic model for atomic electrons energies, for a 270 MeV mediator with $\Gamma=1$ keV,
decaying into 90 MeV final state particles. In this configuration, for the annihilation with at-rest electrons, the positron threshold energy
would be approximately 25 GeV, but thanks to atomic motions also lower-energy positrons can undergo this reaction.

\section{Conclusion}

 The package DMG4 for the simulation of light dark matter production in fixed target experiments is created. It can be
used in full simulation programs based on the Geant4 framework. The subpackage DarkMatter containing a collection of cross sections
is weakly connected to Geant4, thus can be used in programs based on other frameworks. The package is widely used
in the NA64 experiment.

 In this paper, numerous developments of the the package are presented, in particular the WW approximation cross sections for the
muon beam \cite{Sieber:2023nkq}, models with semivisible $A^\prime$ (inelastic Dark Matter) \cite{Filimonova_2022} and several
important improvements for the annihilation processes.

 The package is available at http://mkirsano.web.cern.ch/mkirsano/DMG4.tar.gz and
 https://gitlab.cern.ch/P348/DMG4 (internal NA64 repository). It is recommended also to contact the
 corresponding author Mikhail Kirsanov about the usage.

\section{Acknowledgements}

 This work was supported by the Ministry of Science and Higher Education (MSHE) and RAS (Russia), MSHE Project FSWW-20230003,
Tomsk Polytechnic University within the assignment of MSHE (Russia), the European Research Council (ERC) under the European Union's Horizon 
2020 research and innovation program (Grant agreement No. 947715 - POKER Starting Grant). This work is partially supported by ICSC -- Centro 
Nazionale di Ricerca in High Performance Computing, Big Data and Quantum Computing, funded by European Union -- NextGenerationEU.
This work is also supported by RyC-030551-I, PID2021-123955NA-100 and CNS2022-135850 funded by MCIN/AEI/FEDER, UE (Spain) and
by ETH Zurich and SNSF Grants No. 186181, 186158, 197346, 216602 (Switzerland).

\section{Appendix A}

In order to check the the procedure ``Maximum-Rejection'' we considered a simple model where there is a pseudo-positron particle moving
forward in a material and loosing a constant, deterministic amount of energy $\alpha$ per unit length. Therefore, if the pseudo-positron pseudo-positron enters in this material at $z=0$ with initial energy $E_0$, the energy at depth $z$ in the material is given by $E(z)=E_0-\alpha z$, for $0<z<E_0/\alpha$. The only other process available for this particle is the annihilation process, with a simplified cross section formula
$\sigma(E) = \sigma_0 \Gamma^2 / ((E-E_R)^2 + \Gamma^2)$.

\begin{figure}[t]
  \centering
  \includegraphics[width=.6\textwidth]{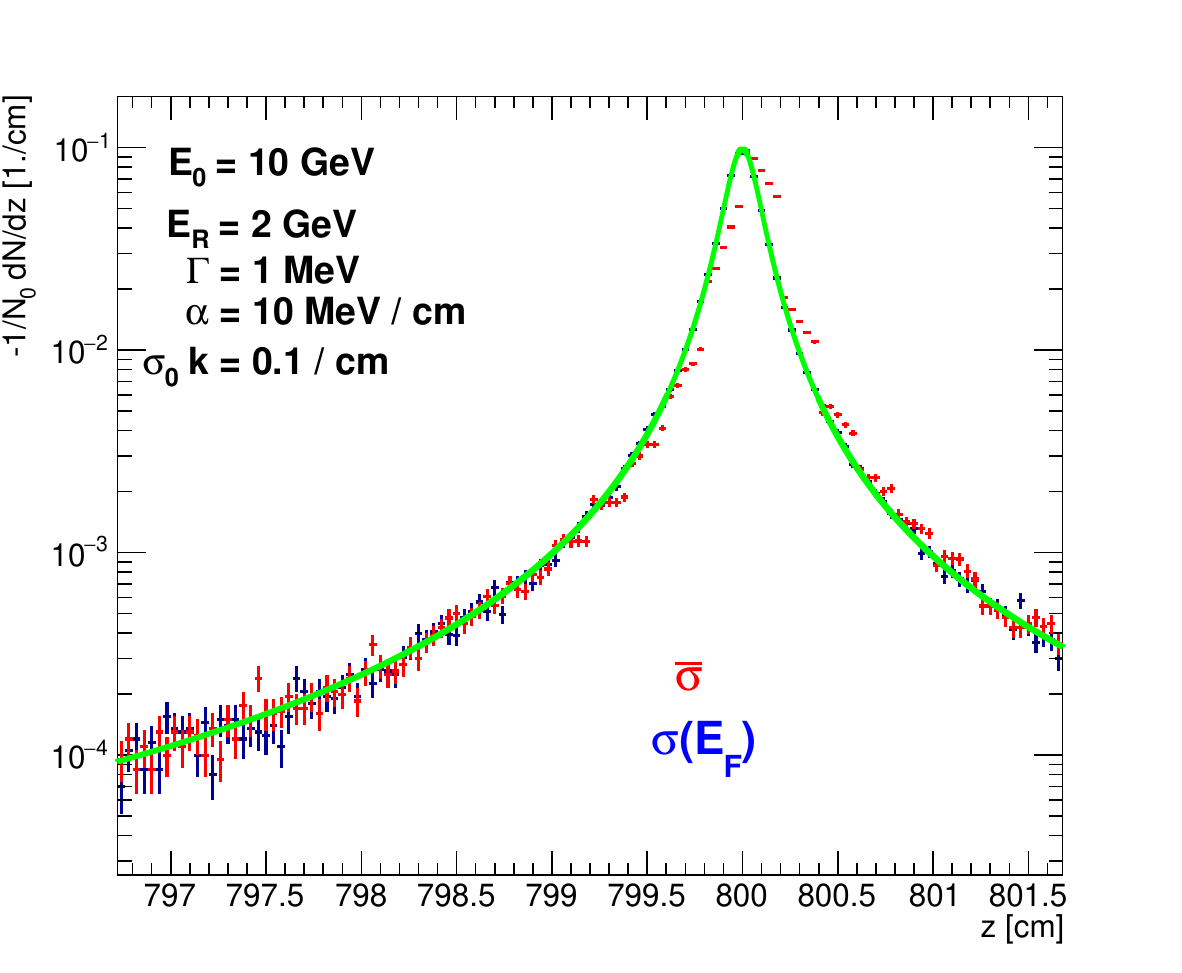}
  \caption{\label{fig:MaximumRejection} The result of checking the ``Maximum-Rejection'' recipe. Green curve - exact solution,
blue points with error bars - using $E_f$ as specified in the recipe, red points with error bars - using mean cross section along
the step instead.}
\end{figure}

If a mono-chromatic beam of pseudo-positrons impinges on the material at $z=0$, the number of particles at a given depth in the material $N(z)$ can be computed from the following equation:
\begin{equation}
  \frac{dN}{dz} = -N(z)k\sigma_0 \frac{\Gamma^2}{(E(z) - E_R)^2 + \Gamma^2}= -N(z)k\sigma_0 \frac{\Gamma^2}{(E_0 - \alpha z - E_R)^2 + \Gamma^2}\; \; ,
 \label{diffeq}
\end{equation}
where $k$ is the number density of electrons per unit volume in the material. This equation can be solved exactly, obtaining:
\begin{equation}
  N(z) = N_0 \cdot e^{-\frac{k\sigma_0 \Gamma}{\alpha} atan \frac{E_0-E_R}{\Gamma}} \cdot
             e^{-\frac{k\sigma_0 \Gamma}{\alpha} atan \frac{E_0-E_R-\alpha z}{\Gamma}}
 \label{diffeqres}
\end{equation}
This gives a number of interactions per unit length, $-\frac{dN}{dz}$, equal to:
\begin{equation}
  -\frac{dN}{dz} = N_0 \cdot e^{-\frac{k\sigma_0 \Gamma}{\alpha} atan \frac{E_0-E_R}{\Gamma}}
	\cdot               e^{-\frac{k\sigma_0 \Gamma}{\alpha} atan \frac{E_0-E_R-\alpha z}{\Gamma}}
          \cdot k\sigma_0 \frac{1}{1 + \left(\frac{E_0-E_R-\alpha z}{\Gamma}\right)^2}.
 \label{diffeqres1}  
\end{equation}
To check the ``Maximum-Rejection'' procedure we made a toy Monte Carlo simulation, accounting for the energy variation across the step as discussed in Sec.~\ref{sec:ene_dep}. Specifically, we implemented a first version computing the reduced probability $p_{red}$ through the cross section at the end of the step, as recommended in the Geant4 manual, and also by using the mean cross section value along the step.
The result is shown in Fig.~\ref{fig:MaximumRejection}. This test confirms the solution suggested by the Geant4 manual.

\clearpage
\bibliographystyle{apsrev4-2}
\bibliography{../Bibliography/bibliographyOther_inspiresFormat,../Bibliography/bibliographyNA64_inspiresFormat,../Bibliography/bibliographyNA64exp_inspiresFormat}

\begin{thebibliography}{39}%
\makeatletter
\providecommand \@ifxundefined [1]{%
 \@ifx{#1\undefined}
}%
\providecommand \@ifnum [1]{%
 \ifnum #1\expandafter \@firstoftwo
 \else \expandafter \@secondoftwo
 \fi
}%
\providecommand \@ifx [1]{%
 \ifx #1\expandafter \@firstoftwo
 \else \expandafter \@secondoftwo
 \fi
}%
\providecommand \natexlab [1]{#1}%
\providecommand \enquote  [1]{``#1''}%
\providecommand \bibnamefont  [1]{#1}%
\providecommand \bibfnamefont [1]{#1}%
\providecommand \citenamefont [1]{#1}%
\providecommand \href@noop [0]{\@secondoftwo}%
\providecommand \href [0]{\begingroup \@sanitize@url \@href}%
\providecommand \@href[1]{\@@startlink{#1}\@@href}%
\providecommand \@@href[1]{\endgroup#1\@@endlink}%
\providecommand \@sanitize@url [0]{\catcode `\\12\catcode `\$12\catcode
  `\&12\catcode `\#12\catcode `\^12\catcode `\_12\catcode `\%12\relax}%
\providecommand \@@startlink[1]{}%
\providecommand \@@endlink[0]{}%
\providecommand \url  [0]{\begingroup\@sanitize@url \@url }%
\providecommand \@url [1]{\endgroup\@href {#1}{\urlprefix }}%
\providecommand \urlprefix  [0]{URL }%
\providecommand \Eprint [0]{\href }%
\providecommand \doibase [0]{https://doi.org/}%
\providecommand \selectlanguage [0]{\@gobble}%
\providecommand \bibinfo  [0]{\@secondoftwo}%
\providecommand \bibfield  [0]{\@secondoftwo}%
\providecommand \translation [1]{[#1]}%
\providecommand \BibitemOpen [0]{}%
\providecommand \bibitemStop [0]{}%
\providecommand \bibitemNoStop [0]{.\EOS\space}%
\providecommand \EOS [0]{\spacefactor3000\relax}%
\providecommand \BibitemShut  [1]{\csname bibitem#1\endcsname}%
\let\auto@bib@innerbib\@empty
\bibitem [{\citenamefont {Bjorken}\ \emph {et~al.}(2009)\citenamefont
  {Bjorken}, \citenamefont {Essig}, \citenamefont {Schuster},\ and\
  \citenamefont {Toro}}]{Bjorken:2009mm}%
  \BibitemOpen
  \bibfield  {author} {\bibinfo {author} {\bibfnamefont {J.~D.}\ \bibnamefont
  {Bjorken}}, \bibinfo {author} {\bibfnamefont {R.}~\bibnamefont {Essig}},
  \bibinfo {author} {\bibfnamefont {P.}~\bibnamefont {Schuster}},\ and\
  \bibinfo {author} {\bibfnamefont {N.}~\bibnamefont {Toro}},\ }\href
  {https://doi.org/10.1103/PhysRevD.80.075018} {\bibfield  {journal} {\bibinfo
  {journal} {Phys. Rev. D}\ }\textbf {\bibinfo {volume} {80}},\ \bibinfo
  {pages} {075018} (\bibinfo {year} {2009})},\ \Eprint
  {https://arxiv.org/abs/0906.0580} {arXiv:0906.0580 [hep-ph]} \BibitemShut
  {NoStop}%
\bibitem [{\citenamefont {Izaguirre}\ \emph {et~al.}(2013)\citenamefont
  {Izaguirre}, \citenamefont {Krnjaic}, \citenamefont {Schuster},\ and\
  \citenamefont {Toro}}]{Izaguirre:2013uxa}%
  \BibitemOpen
  \bibfield  {author} {\bibinfo {author} {\bibfnamefont {E.}~\bibnamefont
  {Izaguirre}}, \bibinfo {author} {\bibfnamefont {G.}~\bibnamefont {Krnjaic}},
  \bibinfo {author} {\bibfnamefont {P.}~\bibnamefont {Schuster}},\ and\
  \bibinfo {author} {\bibfnamefont {N.}~\bibnamefont {Toro}},\ }\href
  {https://doi.org/10.1103/PhysRevD.88.114015} {\bibfield  {journal} {\bibinfo
  {journal} {Phys. Rev. D}\ }\textbf {\bibinfo {volume} {88}},\ \bibinfo
  {pages} {114015} (\bibinfo {year} {2013})},\ \Eprint
  {https://arxiv.org/abs/1307.6554} {arXiv:1307.6554 [hep-ph]} \BibitemShut
  {NoStop}%
\bibitem [{\citenamefont {Batell}\ \emph {et~al.}(2009)\citenamefont {Batell},
  \citenamefont {Pospelov},\ and\ \citenamefont {Ritz}}]{Batell:2009di}%
  \BibitemOpen
  \bibfield  {author} {\bibinfo {author} {\bibfnamefont {B.}~\bibnamefont
  {Batell}}, \bibinfo {author} {\bibfnamefont {M.}~\bibnamefont {Pospelov}},\
  and\ \bibinfo {author} {\bibfnamefont {A.}~\bibnamefont {Ritz}},\ }\href
  {https://doi.org/10.1103/PhysRevD.80.095024} {\bibfield  {journal} {\bibinfo
  {journal} {Phys. Rev. D}\ }\textbf {\bibinfo {volume} {80}},\ \bibinfo
  {pages} {095024} (\bibinfo {year} {2009})},\ \Eprint
  {https://arxiv.org/abs/0906.5614} {arXiv:0906.5614 [hep-ph]} \BibitemShut
  {NoStop}%
\bibitem [{\citenamefont {Izaguirre}\ \emph {et~al.}(2015)\citenamefont
  {Izaguirre}, \citenamefont {Krnjaic}, \citenamefont {Schuster},\ and\
  \citenamefont {Toro}}]{Izaguirre:2014bca}%
  \BibitemOpen
  \bibfield  {author} {\bibinfo {author} {\bibfnamefont {E.}~\bibnamefont
  {Izaguirre}}, \bibinfo {author} {\bibfnamefont {G.}~\bibnamefont {Krnjaic}},
  \bibinfo {author} {\bibfnamefont {P.}~\bibnamefont {Schuster}},\ and\
  \bibinfo {author} {\bibfnamefont {N.}~\bibnamefont {Toro}},\ }\href
  {https://doi.org/10.1103/PhysRevD.91.094026} {\bibfield  {journal} {\bibinfo
  {journal} {Phys. Rev. D}\ }\textbf {\bibinfo {volume} {91}},\ \bibinfo
  {pages} {094026} (\bibinfo {year} {2015})},\ \Eprint
  {https://arxiv.org/abs/1411.1404} {arXiv:1411.1404 [hep-ph]} \BibitemShut
  {NoStop}%
\bibitem [{\citenamefont {Agostinelli}\ \emph {et~al.}(2003)\citenamefont
  {Agostinelli} \emph {et~al.}}]{Agostinelli:2002hh}%
  \BibitemOpen
  \bibfield  {author} {\bibinfo {author} {\bibfnamefont {S.}~\bibnamefont
  {Agostinelli}} \emph {et~al.} (\bibinfo {collaboration} {GEANT4}),\ }\href
  {https://doi.org/10.1016/S0168-9002(03)01368-8} {\bibfield  {journal}
  {\bibinfo  {journal} {Nucl. Instrum. Meth. A}\ }\textbf {\bibinfo {volume}
  {506}},\ \bibinfo {pages} {250} (\bibinfo {year} {2003})}\BibitemShut
  {NoStop}%
\bibitem [{\citenamefont {Bondi}\ \emph {et~al.}(2021)\citenamefont {Bondi},
  \citenamefont {Celentano}, \citenamefont {Dusaev}, \citenamefont
  {Kirpichnikov}, \citenamefont {Kirsanov}, \citenamefont {Krasnikov},
  \citenamefont {Marsicano},\ and\ \citenamefont {Shchukin}}]{Bondi:2021nfp}%
  \BibitemOpen
  \bibfield  {author} {\bibinfo {author} {\bibfnamefont {M.}~\bibnamefont
  {Bondi}}, \bibinfo {author} {\bibfnamefont {A.}~\bibnamefont {Celentano}},
  \bibinfo {author} {\bibfnamefont {R.~R.}\ \bibnamefont {Dusaev}}, \bibinfo
  {author} {\bibfnamefont {D.~V.}\ \bibnamefont {Kirpichnikov}}, \bibinfo
  {author} {\bibfnamefont {M.~M.}\ \bibnamefont {Kirsanov}}, \bibinfo {author}
  {\bibfnamefont {N.~V.}\ \bibnamefont {Krasnikov}}, \bibinfo {author}
  {\bibfnamefont {L.}~\bibnamefont {Marsicano}},\ and\ \bibinfo {author}
  {\bibfnamefont {D.}~\bibnamefont {Shchukin}},\ }\href
  {https://doi.org/10.1016/j.cpc.2021.108129} {\bibfield  {journal} {\bibinfo
  {journal} {Comput. Phys. Commun.}\ }\textbf {\bibinfo {volume} {269}},\
  \bibinfo {pages} {108129} (\bibinfo {year} {2021})},\ \Eprint
  {https://arxiv.org/abs/2101.12192} {arXiv:2101.12192 [hep-ph]} \BibitemShut
  {NoStop}%
\bibitem [{\citenamefont {Holdom}(1986)}]{Holdom:1985ag}%
  \BibitemOpen
  \bibfield  {author} {\bibinfo {author} {\bibfnamefont {B.}~\bibnamefont
  {Holdom}},\ }\href {https://doi.org/10.1016/0370-2693(86)91377-8} {\bibfield
  {journal} {\bibinfo  {journal} {Phys. Lett. B}\ }\textbf {\bibinfo {volume}
  {166}},\ \bibinfo {pages} {196} (\bibinfo {year} {1986})}\BibitemShut
  {NoStop}%
\bibitem [{\citenamefont {Battaglieri}\ \emph {et~al.}(2017)\citenamefont
  {Battaglieri} \emph {et~al.}}]{Battaglieri:2017aum}%
  \BibitemOpen
  \bibfield  {author} {\bibinfo {author} {\bibfnamefont {M.}~\bibnamefont
  {Battaglieri}} \emph {et~al.},\ }in\ \href@noop {} {\emph {\bibinfo
  {booktitle} {{U.S. Cosmic Visions: New Ideas in Dark Matter}}}}\ (\bibinfo
  {year} {2017})\ \Eprint {https://arxiv.org/abs/1707.04591} {arXiv:1707.04591
  [hep-ph]} \BibitemShut {NoStop}%
\bibitem [{\citenamefont {Beacham}\ \emph {et~al.}(2020)\citenamefont {Beacham}
  \emph {et~al.}}]{Beacham:2019nyx}%
  \BibitemOpen
  \bibfield  {author} {\bibinfo {author} {\bibfnamefont {J.}~\bibnamefont
  {Beacham}} \emph {et~al.},\ }\href {https://doi.org/10.1088/1361-6471/ab4cd2}
  {\bibfield  {journal} {\bibinfo  {journal} {J. Phys. G}\ }\textbf {\bibinfo
  {volume} {47}},\ \bibinfo {pages} {010501} (\bibinfo {year} {2020})},\
  \Eprint {https://arxiv.org/abs/1901.09966} {arXiv:1901.09966 [hep-ex]}
  \BibitemShut {NoStop}%
\bibitem [{\citenamefont {{ISO Central Secretary}}(2012)}]{iso_uml2_standard}%
  \BibitemOpen
  \bibfield  {author} {\bibinfo {author} {\bibnamefont {{ISO Central
  Secretary}}},\ }\href {https://www.iso.org/standard/32624.html} {\emph
  {\bibinfo {title} {Information technology -- Object Management Group Unified
  Modeling Language (OMG UML) -- Part 1: Infrastructure}}},\ \bibinfo {type}
  {Standard}\ \bibinfo {number} {ISO/IEC 19505-1:2012}\ (\bibinfo
  {institution} {International Organization for Standardization},\ \bibinfo
  {address} {Geneva, CH},\ \bibinfo {year} {2012})\BibitemShut {NoStop}%
\bibitem [{PDG()}]{PDG_particles}%
  \BibitemOpen
  \href@noop {} {\bibinfo {title} {{Monte Carlo Particle Numbering Scheme}}},\
  \bibinfo {howpublished}
  {\url{https://pdg.lbl.gov/2019/reviews/rpp2019-rev-monte-carlo-numbering.pdf}},\
  \bibinfo {note} {accessed: 2021-01-25}\BibitemShut {NoStop}%
\bibitem [{\citenamefont {Tucker-Smith}\ and\ \citenamefont
  {Weiner}(2001)}]{Tucker-Smith:2001myb}%
  \BibitemOpen
  \bibfield  {author} {\bibinfo {author} {\bibfnamefont {D.}~\bibnamefont
  {Tucker-Smith}}\ and\ \bibinfo {author} {\bibfnamefont {N.}~\bibnamefont
  {Weiner}},\ }\href {https://doi.org/10.1103/PhysRevD.64.043502} {\bibfield
  {journal} {\bibinfo  {journal} {Phys. Rev. D}\ }\textbf {\bibinfo {volume}
  {64}},\ \bibinfo {pages} {043502} (\bibinfo {year} {2001})},\ \Eprint
  {https://arxiv.org/abs/hep-ph/0101138} {arXiv:hep-ph/0101138} \BibitemShut
  {NoStop}%
\bibitem [{\citenamefont {Dusaev}\ \emph {et~al.}(2020)\citenamefont {Dusaev},
  \citenamefont {Kirpichnikov},\ and\ \citenamefont
  {Kirsanov}}]{Dusaev:2020gxi}%
  \BibitemOpen
  \bibfield  {author} {\bibinfo {author} {\bibfnamefont {R.~R.}\ \bibnamefont
  {Dusaev}}, \bibinfo {author} {\bibfnamefont {D.~V.}\ \bibnamefont
  {Kirpichnikov}},\ and\ \bibinfo {author} {\bibfnamefont {M.~M.}\ \bibnamefont
  {Kirsanov}},\ }\href {https://doi.org/10.1103/PhysRevD.102.055018} {\bibfield
   {journal} {\bibinfo  {journal} {Phys. Rev. D}\ }\textbf {\bibinfo {volume}
  {102}},\ \bibinfo {pages} {055018} (\bibinfo {year} {2020})},\ \Eprint
  {https://arxiv.org/abs/2004.04469} {arXiv:2004.04469 [hep-ph]} \BibitemShut
  {NoStop}%
\bibitem [{\citenamefont {Marsicano}\ \emph {et~al.}(2018)\citenamefont
  {Marsicano}, \citenamefont {Battaglieri}, \citenamefont {Bond\'\i{}},
  \citenamefont {Carvajal}, \citenamefont {Celentano}, \citenamefont
  {De~Napoli}, \citenamefont {De~Vita}, \citenamefont {Nardi}, \citenamefont
  {Raggi},\ and\ \citenamefont {Valente}}]{Marsicano:2018glj}%
  \BibitemOpen
  \bibfield  {author} {\bibinfo {author} {\bibfnamefont {L.}~\bibnamefont
  {Marsicano}}, \bibinfo {author} {\bibfnamefont {M.}~\bibnamefont
  {Battaglieri}}, \bibinfo {author} {\bibfnamefont {M.}~\bibnamefont
  {Bond\'\i{}}}, \bibinfo {author} {\bibfnamefont {C.~D.~R.}\ \bibnamefont
  {Carvajal}}, \bibinfo {author} {\bibfnamefont {A.}~\bibnamefont {Celentano}},
  \bibinfo {author} {\bibfnamefont {M.}~\bibnamefont {De~Napoli}}, \bibinfo
  {author} {\bibfnamefont {R.}~\bibnamefont {De~Vita}}, \bibinfo {author}
  {\bibfnamefont {E.}~\bibnamefont {Nardi}}, \bibinfo {author} {\bibfnamefont
  {M.}~\bibnamefont {Raggi}},\ and\ \bibinfo {author} {\bibfnamefont
  {P.}~\bibnamefont {Valente}},\ }\href
  {https://doi.org/10.1103/PhysRevLett.121.041802} {\bibfield  {journal}
  {\bibinfo  {journal} {Phys. Rev. Lett.}\ }\textbf {\bibinfo {volume} {121}},\
  \bibinfo {pages} {041802} (\bibinfo {year} {2018})},\ \Eprint
  {https://arxiv.org/abs/1807.05884} {arXiv:1807.05884 [hep-ex]} \BibitemShut
  {NoStop}%
\bibitem [{\citenamefont {Voronchikhin}\ and\ \citenamefont
  {Kirpichnikov}(2022)}]{Voronchikhin_2022}%
  \BibitemOpen
  \bibfield  {author} {\bibinfo {author} {\bibfnamefont {I.~V.}\ \bibnamefont
  {Voronchikhin}}\ and\ \bibinfo {author} {\bibfnamefont {D.~V.}\ \bibnamefont
  {Kirpichnikov}},\ }\bibfield  {journal} {\bibinfo  {journal} {Physical Review
  D}\ }\textbf {\bibinfo {volume} {106}},\ \href
  {https://doi.org/10.1103/physrevd.106.115041} {10.1103/physrevd.106.115041}
  (\bibinfo {year} {2022})\BibitemShut {NoStop}%
\bibitem [{\citenamefont {Lee}\ \emph {et~al.}(2014)\citenamefont {Lee},
  \citenamefont {Park},\ and\ \citenamefont {Sanz}}]{lee:2014GMDM}%
  \BibitemOpen
  \bibfield  {author} {\bibinfo {author} {\bibfnamefont {H.~M.}\ \bibnamefont
  {Lee}}, \bibinfo {author} {\bibfnamefont {M.}~\bibnamefont {Park}},\ and\
  \bibinfo {author} {\bibfnamefont {V.}~\bibnamefont {Sanz}},\ }\href
  {https://doi.org/10.1140/epjc/s10052-014-2715-8} {\bibfield  {journal}
  {\bibinfo  {journal} {Eur. Phys. J. C}\ }\textbf {\bibinfo {volume} {74}},\
  \bibinfo {pages} {2715} (\bibinfo {year} {2014})},\ \Eprint
  {https://arxiv.org/abs/1306.4107} {arXiv:1306.4107 [hep-ph]} \BibitemShut
  {NoStop}%
\bibitem [{\citenamefont {Kang}\ and\ \citenamefont
  {Lee}(2020)}]{Kang:2020-LGMDM}%
  \BibitemOpen
  \bibfield  {author} {\bibinfo {author} {\bibfnamefont {Y.-J.}\ \bibnamefont
  {Kang}}\ and\ \bibinfo {author} {\bibfnamefont {H.~M.}\ \bibnamefont {Lee}},\
  }\bibfield  {journal} {\bibinfo  {journal} {The European Physical Journal C}\
  }\textbf {\bibinfo {volume} {80}},\ \href
  {https://doi.org/10.1140/epjc/s10052-020-8153-x}
  {10.1140/epjc/s10052-020-8153-x} (\bibinfo {year} {2020}),\ \Eprint
  {https://arxiv.org/abs/2001.04868} {arXiv:2001.04868} \BibitemShut {NoStop}%
\bibitem [{\citenamefont {Filimonova}\ \emph {et~al.}(2022)\citenamefont
  {Filimonova}, \citenamefont {Junius}, \citenamefont {Honorez},\ and\
  \citenamefont {Westhoff}}]{Filimonova_2022}%
  \BibitemOpen
  \bibfield  {author} {\bibinfo {author} {\bibfnamefont {A.}~\bibnamefont
  {Filimonova}}, \bibinfo {author} {\bibfnamefont {S.}~\bibnamefont {Junius}},
  \bibinfo {author} {\bibfnamefont {L.~L.}\ \bibnamefont {Honorez}},\ and\
  \bibinfo {author} {\bibfnamefont {S.}~\bibnamefont {Westhoff}},\ }\bibfield
  {journal} {\bibinfo  {journal} {Journal of High Energy Physics}\ }\textbf
  {\bibinfo {volume} {2022}},\ \href {https://doi.org/10.1007/jhep06(2022)048}
  {10.1007/jhep06(2022)048} (\bibinfo {year} {2022})\BibitemShut {NoStop}%
\bibitem [{\citenamefont {Mohlabeng}(2019)}]{Mohlabeng:2019vrz}%
  \BibitemOpen
  \bibfield  {author} {\bibinfo {author} {\bibfnamefont {G.}~\bibnamefont
  {Mohlabeng}},\ }\href {https://doi.org/10.1103/PhysRevD.99.115001} {\bibfield
   {journal} {\bibinfo  {journal} {Phys. Rev. D}\ }\textbf {\bibinfo {volume}
  {99}},\ \bibinfo {pages} {115001} (\bibinfo {year} {2019})},\ \Eprint
  {https://arxiv.org/abs/1902.05075} {arXiv:1902.05075 [hep-ph]} \BibitemShut
  {NoStop}%
\bibitem [{\citenamefont {Cazzaniga}\ \emph {et~al.}(2021)\citenamefont
  {Cazzaniga} \emph {et~al.}}]{NA64:2021acr}%
  \BibitemOpen
  \bibfield  {author} {\bibinfo {author} {\bibfnamefont {C.}~\bibnamefont
  {Cazzaniga}} \emph {et~al.} (\bibinfo {collaboration} {NA64}),\ }\href
  {https://doi.org/10.1140/epjc/s10052-021-09705-5} {\bibfield  {journal}
  {\bibinfo  {journal} {Eur. Phys. J. C}\ }\textbf {\bibinfo {volume} {81}},\
  \bibinfo {pages} {959} (\bibinfo {year} {2021})},\ \Eprint
  {https://arxiv.org/abs/2107.02021} {arXiv:2107.02021 [hep-ex]} \BibitemShut
  {NoStop}%
\bibitem [{\citenamefont {He}\ \emph {et~al.}(1991{\natexlab{a}})\citenamefont
  {He}, \citenamefont {Joshi}, \citenamefont {Lew},\ and\ \citenamefont
  {Volkas}}]{He:1990pn}%
  \BibitemOpen
  \bibfield  {author} {\bibinfo {author} {\bibfnamefont {X.~G.}\ \bibnamefont
  {He}}, \bibinfo {author} {\bibfnamefont {G.~C.}\ \bibnamefont {Joshi}},
  \bibinfo {author} {\bibfnamefont {H.}~\bibnamefont {Lew}},\ and\ \bibinfo
  {author} {\bibfnamefont {R.~R.}\ \bibnamefont {Volkas}},\ }\href
  {https://doi.org/10.1103/PhysRevD.43.R22} {\bibfield  {journal} {\bibinfo
  {journal} {Phys. Rev. D}\ }\textbf {\bibinfo {volume} {43}},\ \bibinfo
  {pages} {22} (\bibinfo {year} {1991}{\natexlab{a}})}\BibitemShut {NoStop}%
\bibitem [{\citenamefont {He}\ \emph {et~al.}(1991{\natexlab{b}})\citenamefont
  {He}, \citenamefont {Joshi}, \citenamefont {Lew},\ and\ \citenamefont
  {Volkas}}]{He:1991qd}%
  \BibitemOpen
  \bibfield  {author} {\bibinfo {author} {\bibfnamefont {X.-G.}\ \bibnamefont
  {He}}, \bibinfo {author} {\bibfnamefont {G.~C.}\ \bibnamefont {Joshi}},
  \bibinfo {author} {\bibfnamefont {H.}~\bibnamefont {Lew}},\ and\ \bibinfo
  {author} {\bibfnamefont {R.~R.}\ \bibnamefont {Volkas}},\ }\href
  {https://doi.org/10.1103/PhysRevD.44.2118} {\bibfield  {journal} {\bibinfo
  {journal} {Phys. Rev. D}\ }\textbf {\bibinfo {volume} {44}},\ \bibinfo
  {pages} {2118} (\bibinfo {year} {1991}{\natexlab{b}})}\BibitemShut {NoStop}%
\bibitem [{\citenamefont {Foot}\ \emph {et~al.}(1994)\citenamefont {Foot},
  \citenamefont {He}, \citenamefont {Lew},\ and\ \citenamefont
  {Volkas}}]{Foot:1994vd}%
  \BibitemOpen
  \bibfield  {author} {\bibinfo {author} {\bibfnamefont {R.}~\bibnamefont
  {Foot}}, \bibinfo {author} {\bibfnamefont {X.~G.}\ \bibnamefont {He}},
  \bibinfo {author} {\bibfnamefont {H.}~\bibnamefont {Lew}},\ and\ \bibinfo
  {author} {\bibfnamefont {R.~R.}\ \bibnamefont {Volkas}},\ }\href
  {https://doi.org/10.1103/PhysRevD.50.4571} {\bibfield  {journal} {\bibinfo
  {journal} {Phys. Rev. D}\ }\textbf {\bibinfo {volume} {50}},\ \bibinfo
  {pages} {4571} (\bibinfo {year} {1994})},\ \Eprint
  {https://arxiv.org/abs/hep-ph/9401250} {arXiv:hep-ph/9401250} \BibitemShut
  {NoStop}%
\bibitem [{\citenamefont {Altmannshofer}\ \emph {et~al.}(2016)\citenamefont
  {Altmannshofer}, \citenamefont {Gori}, \citenamefont {Profumo},\ and\
  \citenamefont {Queiroz}}]{Altmannshofer:2016jzy}%
  \BibitemOpen
  \bibfield  {author} {\bibinfo {author} {\bibfnamefont {W.}~\bibnamefont
  {Altmannshofer}}, \bibinfo {author} {\bibfnamefont {S.}~\bibnamefont {Gori}},
  \bibinfo {author} {\bibfnamefont {S.}~\bibnamefont {Profumo}},\ and\ \bibinfo
  {author} {\bibfnamefont {F.~S.}\ \bibnamefont {Queiroz}},\ }\href
  {https://doi.org/10.1007/JHEP12(2016)106} {\bibfield  {journal} {\bibinfo
  {journal} {JHEP}\ }\textbf {\bibinfo {volume} {12}},\ \bibinfo {pages}
  {106}},\ \Eprint {https://arxiv.org/abs/1609.04026} {arXiv:1609.04026
  [hep-ph]} \BibitemShut {NoStop}%
\bibitem [{\citenamefont {Kile}\ \emph {et~al.}(2015)\citenamefont {Kile},
  \citenamefont {Kobach},\ and\ \citenamefont {Soni}}]{Kile:2014jea}%
  \BibitemOpen
  \bibfield  {author} {\bibinfo {author} {\bibfnamefont {J.}~\bibnamefont
  {Kile}}, \bibinfo {author} {\bibfnamefont {A.}~\bibnamefont {Kobach}},\ and\
  \bibinfo {author} {\bibfnamefont {A.}~\bibnamefont {Soni}},\ }\href
  {https://doi.org/10.1016/j.physletb.2015.04.005} {\bibfield  {journal}
  {\bibinfo  {journal} {Phys. Lett. B}\ }\textbf {\bibinfo {volume} {744}},\
  \bibinfo {pages} {330} (\bibinfo {year} {2015})},\ \Eprint
  {https://arxiv.org/abs/1411.1407} {arXiv:1411.1407 [hep-ph]} \BibitemShut
  {NoStop}%
\bibitem [{\citenamefont {Park}\ \emph {et~al.}(2016)\citenamefont {Park},
  \citenamefont {Kim},\ and\ \citenamefont {Park}}]{Park:2015gdo}%
  \BibitemOpen
  \bibfield  {author} {\bibinfo {author} {\bibfnamefont {J.-C.}\ \bibnamefont
  {Park}}, \bibinfo {author} {\bibfnamefont {J.}~\bibnamefont {Kim}},\ and\
  \bibinfo {author} {\bibfnamefont {S.~C.}\ \bibnamefont {Park}},\ }\href
  {https://doi.org/10.1016/j.physletb.2015.11.035} {\bibfield  {journal}
  {\bibinfo  {journal} {Phys. Lett. B}\ }\textbf {\bibinfo {volume} {752}},\
  \bibinfo {pages} {59} (\bibinfo {year} {2016})},\ \Eprint
  {https://arxiv.org/abs/1505.04620} {arXiv:1505.04620 [hep-ph]} \BibitemShut
  {NoStop}%
\bibitem [{\citenamefont {Liu}\ \emph {et~al.}(2017)\citenamefont {Liu},
  \citenamefont {McKeen},\ and\ \citenamefont {Miller}}]{Liu:2016mqv}%
  \BibitemOpen
  \bibfield  {author} {\bibinfo {author} {\bibfnamefont {Y.-S.}\ \bibnamefont
  {Liu}}, \bibinfo {author} {\bibfnamefont {D.}~\bibnamefont {McKeen}},\ and\
  \bibinfo {author} {\bibfnamefont {G.~A.}\ \bibnamefont {Miller}},\ }\href
  {https://doi.org/10.1103/PhysRevD.95.036010} {\bibfield  {journal} {\bibinfo
  {journal} {Phys. Rev. D}\ }\textbf {\bibinfo {volume} {95}},\ \bibinfo
  {pages} {036010} (\bibinfo {year} {2017})},\ \Eprint
  {https://arxiv.org/abs/1609.06781} {arXiv:1609.06781 [hep-ph]} \BibitemShut
  {NoStop}%
\bibitem [{\citenamefont {Liu}\ and\ \citenamefont
  {Miller}(2017)}]{Liu:2017htz}%
  \BibitemOpen
  \bibfield  {author} {\bibinfo {author} {\bibfnamefont {Y.-S.}\ \bibnamefont
  {Liu}}\ and\ \bibinfo {author} {\bibfnamefont {G.~A.}\ \bibnamefont
  {Miller}},\ }\href {https://doi.org/10.1103/PhysRevD.96.016004} {\bibfield
  {journal} {\bibinfo  {journal} {Phys. Rev. D}\ }\textbf {\bibinfo {volume}
  {96}},\ \bibinfo {pages} {016004} (\bibinfo {year} {2017})},\ \Eprint
  {https://arxiv.org/abs/1705.01633} {arXiv:1705.01633 [hep-ph]} \BibitemShut
  {NoStop}%
\bibitem [{\citenamefont {Kirpichnikov}\ \emph {et~al.}(2021)\citenamefont
  {Kirpichnikov}, \citenamefont {Sieber}, \citenamefont {Bueno}, \citenamefont
  {Crivelli},\ and\ \citenamefont {Kirsanov}}]{Kirpichnikov:2021jev}%
  \BibitemOpen
  \bibfield  {author} {\bibinfo {author} {\bibfnamefont {D.~V.}\ \bibnamefont
  {Kirpichnikov}}, \bibinfo {author} {\bibfnamefont {H.}~\bibnamefont
  {Sieber}}, \bibinfo {author} {\bibfnamefont {L.~M.}\ \bibnamefont {Bueno}},
  \bibinfo {author} {\bibfnamefont {P.}~\bibnamefont {Crivelli}},\ and\
  \bibinfo {author} {\bibfnamefont {M.~M.}\ \bibnamefont {Kirsanov}},\ }\href
  {https://doi.org/10.1103/PhysRevD.104.076012} {\bibfield  {journal} {\bibinfo
   {journal} {Phys. Rev. D}\ }\textbf {\bibinfo {volume} {104}},\ \bibinfo
  {pages} {076012} (\bibinfo {year} {2021})},\ \Eprint
  {https://arxiv.org/abs/2107.13297} {arXiv:2107.13297 [hep-ph]} \BibitemShut
  {NoStop}%
\bibitem [{\citenamefont {Sieber}\ \emph {et~al.}(2023)\citenamefont {Sieber},
  \citenamefont {Kirpichnikov}, \citenamefont {Voronchikhin}, \citenamefont
  {Crivelli}, \citenamefont {Gninenko}, \citenamefont {Kirsanov}, \citenamefont
  {Krasnikov}, \citenamefont {Molina-Bueno},\ and\ \citenamefont
  {Sekatskii}}]{Sieber:2023nkq}%
  \BibitemOpen
  \bibfield  {author} {\bibinfo {author} {\bibfnamefont {H.}~\bibnamefont
  {Sieber}}, \bibinfo {author} {\bibfnamefont {D.~V.}\ \bibnamefont
  {Kirpichnikov}}, \bibinfo {author} {\bibfnamefont {I.~V.}\ \bibnamefont
  {Voronchikhin}}, \bibinfo {author} {\bibfnamefont {P.}~\bibnamefont
  {Crivelli}}, \bibinfo {author} {\bibfnamefont {S.~N.}\ \bibnamefont
  {Gninenko}}, \bibinfo {author} {\bibfnamefont {M.~M.}\ \bibnamefont
  {Kirsanov}}, \bibinfo {author} {\bibfnamefont {N.~V.}\ \bibnamefont
  {Krasnikov}}, \bibinfo {author} {\bibfnamefont {L.}~\bibnamefont
  {Molina-Bueno}},\ and\ \bibinfo {author} {\bibfnamefont {S.~K.}\ \bibnamefont
  {Sekatskii}},\ }\href {https://doi.org/10.1103/PhysRevD.108.056018}
  {\bibfield  {journal} {\bibinfo  {journal} {Phys. Rev. D}\ }\textbf {\bibinfo
  {volume} {108}},\ \bibinfo {pages} {056018} (\bibinfo {year} {2023})},\
  \Eprint {https://arxiv.org/abs/2305.09015} {arXiv:2305.09015 [hep-ph]}
  \BibitemShut {NoStop}%
\bibitem [{\citenamefont {Bogdanov}\ \emph {et~al.}(2006)\citenamefont
  {Bogdanov}, \citenamefont {Burkhardt}, \citenamefont {Ivanchenko},
  \citenamefont {Kelner}, \citenamefont {Kokoulin}, \citenamefont {Maire},
  \citenamefont {Rybin},\ and\ \citenamefont {Urban}}]{Bogdanov:2006kr}%
  \BibitemOpen
  \bibfield  {author} {\bibinfo {author} {\bibfnamefont {A.~G.}\ \bibnamefont
  {Bogdanov}}, \bibinfo {author} {\bibfnamefont {H.}~\bibnamefont {Burkhardt}},
  \bibinfo {author} {\bibfnamefont {V.~N.}\ \bibnamefont {Ivanchenko}},
  \bibinfo {author} {\bibfnamefont {S.~R.}\ \bibnamefont {Kelner}}, \bibinfo
  {author} {\bibfnamefont {R.~P.}\ \bibnamefont {Kokoulin}}, \bibinfo {author}
  {\bibfnamefont {M.}~\bibnamefont {Maire}}, \bibinfo {author} {\bibfnamefont
  {A.~M.}\ \bibnamefont {Rybin}},\ and\ \bibinfo {author} {\bibfnamefont
  {L.}~\bibnamefont {Urban}},\ }\href {https://doi.org/10.1109/TNS.2006.872633}
  {\bibfield  {journal} {\bibinfo  {journal} {IEEE Trans. Nucl. Sci.}\ }\textbf
  {\bibinfo {volume} {53}},\ \bibinfo {pages} {513} (\bibinfo {year}
  {2006})}\BibitemShut {NoStop}%
\bibitem [{\citenamefont {Kolmogorov}(1933)}]{Kolmogorov:1933sde}%
  \BibitemOpen
  \bibfield  {author} {\bibinfo {author} {\bibfnamefont {A.}~\bibnamefont
  {Kolmogorov}},\ }\href@noop {} {\bibfield  {journal} {\bibinfo  {journal} {G.
  Ist. Ital. Attuari.}\ }\textbf {\bibinfo {volume} {4}},\ \bibinfo {pages}
  {83} (\bibinfo {year} {1933})}\BibitemShut {NoStop}%
\bibitem [{\citenamefont {Smirnov}(1948)}]{Smirnov:1948tfe}%
  \BibitemOpen
  \bibfield  {author} {\bibinfo {author} {\bibfnamefont {N.}~\bibnamefont
  {Smirnov}},\ }\href {https://doi.org/10.1214/aoms/1177730256} {\bibfield
  {journal} {\bibinfo  {journal} {Ann. Math. Statist.}\ }\textbf {\bibinfo
  {volume} {19}},\ \bibinfo {pages} {279} (\bibinfo {year} {1948})}\BibitemShut
  {NoStop}%
\bibitem [{\citenamefont {Andreev}\ \emph
  {et~al.}(2022{\natexlab{a}})\citenamefont {Andreev} \emph
  {et~al.}}]{NA64:2022rme}%
  \BibitemOpen
  \bibfield  {author} {\bibinfo {author} {\bibfnamefont {Y.~M.}\ \bibnamefont
  {Andreev}} \emph {et~al.} (\bibinfo {collaboration} {NA64}),\ }\href
  {https://doi.org/10.1103/PhysRevD.106.032015} {\bibfield  {journal} {\bibinfo
   {journal} {Phys. Rev. D}\ }\textbf {\bibinfo {volume} {106}},\ \bibinfo
  {pages} {032015} (\bibinfo {year} {2022}{\natexlab{a}})},\ \Eprint
  {https://arxiv.org/abs/2206.03101} {arXiv:2206.03101 [hep-ex]} \BibitemShut
  {NoStop}%
\bibitem [{\citenamefont {Andreev}\ \emph
  {et~al.}(2022{\natexlab{b}})\citenamefont {Andreev} \emph
  {et~al.}}]{NA64:2022yly}%
  \BibitemOpen
  \bibfield  {author} {\bibinfo {author} {\bibfnamefont {Y.~M.}\ \bibnamefont
  {Andreev}} \emph {et~al.} (\bibinfo {collaboration} {NA64}),\ }\href
  {https://doi.org/10.1103/PhysRevLett.129.161801} {\bibfield  {journal}
  {\bibinfo  {journal} {Phys. Rev. Lett.}\ }\textbf {\bibinfo {volume} {129}},\
  \bibinfo {pages} {161801} (\bibinfo {year} {2022}{\natexlab{b}})},\ \Eprint
  {https://arxiv.org/abs/2207.09979} {arXiv:2207.09979 [hep-ex]} \BibitemShut
  {NoStop}%
\bibitem [{Note1()}]{Note1}%
  \BibitemOpen
  \bibinfo {note} {Incidentally, we note that in the Geant4 physics reference
  manual for release 11.2, rev 8.0, in Sec. 7.4 \protect \textit {``Correcting
  the Cross Section for Energy Variation''} the prescription to use of $\sigma
  (E_f)$ in the computation of $p_{red}$ is supported by a reference to ~\cite
  {Ivanchenko:1991me}. However, in that work no justification of this method is
  provided.}\BibitemShut {Stop}%
\bibitem [{Note2()}]{Note2}%
  \BibitemOpen
  \bibinfo {note} {We incidentally note that this model is the same adopted by
  the Geant4 software to handle atomic motion effects in Compton
  scattering.}\BibitemShut {Stop}%
\bibitem [{Note3()}]{Note3}%
  \BibitemOpen
  \bibinfo {note} {The extension of the code to handle final state with
  particles having different masses is foreseen for the next DMG4
  release.}\BibitemShut {Stop}%
\bibitem [{\citenamefont {Ivanchenko}\ \emph {et~al.}(1991)\citenamefont
  {Ivanchenko}, \citenamefont {Bukin}, \citenamefont {Dubrovin}, \citenamefont
  {Eidelman}, \citenamefont {Grozina},\ and\ \citenamefont
  {Tayursky}}]{Ivanchenko:1991me}%
  \BibitemOpen
  \bibfield  {author} {\bibinfo {author} {\bibfnamefont {V.~N.}\ \bibnamefont
  {Ivanchenko}}, \bibinfo {author} {\bibfnamefont {A.~D.}\ \bibnamefont
  {Bukin}}, \bibinfo {author} {\bibfnamefont {M.~S.}\ \bibnamefont {Dubrovin}},
  \bibinfo {author} {\bibfnamefont {S.~I.}\ \bibnamefont {Eidelman}}, \bibinfo
  {author} {\bibfnamefont {N.~A.}\ \bibnamefont {Grozina}},\ and\ \bibinfo
  {author} {\bibfnamefont {V.~A.}\ \bibnamefont {Tayursky}},\ }in\ \href@noop
  {} {\emph {\bibinfo {booktitle} {{Workshop on Detector and Event Simulation
  in High-energy Physics (MC '91)}}}}\ (\bibinfo {year} {1991})\ pp.\ \bibinfo
  {pages} {79--85}\BibitemShut {NoStop}%
\end{thebibliography}%

\end{document}